\documentclass[preprint,aps,prd,superscriptaddress,preprintnumbers,letterpaper,natbib,floatfix,nofootinbib]{revtex4-1}

\usepackage{comment}
\usepackage{graphicx} 
\usepackage{amsmath}
\usepackage{amssymb}
\usepackage{bbold}
\usepackage{caption}
\usepackage{subcaption}
\usepackage{float}

\usepackage{caption}
\usepackage{tikz,xcolor,hyperref}

\definecolor{lime}{HTML}{A6CE39}
\DeclareRobustCommand{\orcidicon}{\hspace{-1mm}
	\begin{tikzpicture}
		\draw[lime, fill=lime] (0,0) 
		circle [radius=0.16] 
		node[white] {{\fontfamily{qag}\selectfont \tiny \,ID}};
		\draw[white, fill=white] (-0.0525,0.095) 
		circle [radius=0.007];
	\end{tikzpicture}
	\hspace{-3mm}
}

\foreach \x in {A, ..., Z}{\expandafter\xdef\csname orcid\x\endcsname{\noexpand\href{https://orcid.org/\csname orcidauthor\x\endcsname}
		{\noexpand\orcidicon}}
}

\newcommand{\sect}[1]{\mathbb{#1}}

\begin{document}

\title{\large Singlet-doublet dark matter beyond freeze-out}
\author{Prudhvi N.~Bhattiprolu\orcidA}
\email{prudhvi.bhattiprolu@fi.infn.it}
\affiliation{INFN, Sezione di Firenze, Via G. Sansone 1, 50019 Sesto Fiorentino, Italy}
\author{Aaron Pierce\orcidB}
\email{atpierce@umich.edu}
\affiliation{Leinweber Center for Theoretical Physics, Department of Physics,\\ University of Michigan, Ann Arbor, MI 48109, USA}
\author{Alexander Takla\orcidC}
\email{atakla@umich.edu}
\affiliation{Leinweber Center for Theoretical Physics, Department of Physics,\\ University of Michigan, Ann Arbor, MI 48109, USA}

\begin{abstract}
The Singlet-Doublet fermion model of dark matter is an economical weak-scale dark matter model that realizes the dark matter abundance through interactions with the electroweak bosons of the Standard Model.  Depending on the size of the Yukawa couplings in the model, the dark matter relic abundance can be produced via freeze-out (including co-annihilation), co-scattering, freeze-in, or the SuperWIMP mechanism.  We analyze the ways this model can realize the dark matter density with emphasis on the small Yukawa coupling regime. In this limit, direct and indirect detection are difficult, but a rich collider phenomenology is possible.
\end{abstract}

\maketitle
\preprint{LITP-26-15}
\tableofcontents
\newpage

\section{Introduction}\label{sec:intro}

Weak-scale dark matter is an attractive dark matter candidate.  The most common realization is the WIMP (Weakly Interacting Massive Particle) paradigm, wherein the dark matter realizes its relic abundance through weak-scale freeze-out. However, weak-scale dark matter can be produced via a wide variety of additional mechanisms, including freeze-out with co-annihilations \cite{Griest:1990kh}, co-scattering \cite{Garny:2017rxs,DAgnolo:2017dbv}, freeze-in \cite{Hall:2009bx}, and the so-called SuperWIMP mechanism \cite{Feng:2003xh}.  These  mechanisms are of increasing interest because they give rise to dark matter candidates that have small direct detection cross sections and can therefore evade the increasingly stringent bounds from these experiments \cite{LZ:2024zvo}.  

Interestingly, all these mechanisms can be realized in different regions of the parameter space for a single, economical model, the Singlet-Doublet Majorana fermion model of dark matter \cite{Arkani-Hamed:2005zuc,Mahbubani:2005pt,DEramo:2007anh,Cohen:2011ec}. For additional studies, see \cite{Cheung:2013dua,Kearney:2016rng, Enberg:2007rp, Abe:2014gua, Calibbi:2015nha, Freitas:2015hsa,Egana-Ugrinovic:2017jib, LopezHonorez:2017zrd, Esch:2018ccs, Arcadi:2018pfo}.  This model has also been of interest because it can replicate the dark matter phenomenology of Split Supersymmetry \cite{Arkani-Hamed:2004ymt, Giudice:2004tc,Pierce:2004mk} for some choices of parameters.  Moreover, the introduction of the $SU(2)$-doublet fermions substantially improves the quality of gauge coupling unification with respect to the Standard Model (SM) \cite{Arkani-Hamed:2005zuc}, and so embedding this Singlet-Doublet model in a grand unified theory (GUT) might be possible \cite{Mahbubani:2005pt}. For reviews that discuss the Singlet-Doublet model among a range of other WIMP possibilities, see \cite{Arcadi:2024ukq, Arcadi:2019lka, Arcadi:2017kky, Cirelli:2024ssz}, also \cite{Paul:2024prs}.

In this paper, we study this dark matter model in detail, with particular focus on cases where co-scattering, freeze-in and the SuperWIMP mechanism are relevant.  Recent work \cite{Bhattiprolu:2025beq} discussed this model in the regime where co-annihilation-assisted freeze-out was the dominant dark matter production mechanism.  A previous discussion of this model in the freeze-in regime \cite{Calibbi:2018fqf} focused on much lighter {$\mathcal O(\rm{keV})$} dark matter; here we consider weak-scale dark matter. Co-annihilation and co-scattering have also been recently analyzed in this model in Ref.~\cite{Paul_2026}. That study was restricted to a slice of parameter space along the so-called $Z$-blind spot where the two Yukawa couplings are equal.\footnote{While we have qualitative agreement with many results in that work, it shows parameter space with dark matter masses well in excess of those expected for a pure doublet, i.e. 1.1 TeV, which we do not find. So we have quantitative disagreement, even in the parameter space of overlap.}  A closely related model where the dark sector is comprised of Dirac fermions was studied, e.g., in \cite{Yaguna:2015mva,Paul:2024prs}

When couplings within a dark sector are small, parts of the dark sector can fall out of chemical or kinetic equilibrium. Our preliminary study, to be released in a companion paper \cite{Bhattiprolu:ToAppear}, suggests that early kinetic decoupling has minimal impact on the relic abundance, a few percent at most. Departures from chemical equilibrium within the dark sector, however, can have a much larger effect and will be a focus of this paper. In this regime, the dark matter abundance must be evolved separately from the other dark-sector abundances; otherwise, the relic abundance can be underestimated.

The structure of this paper is as follows.  In Sec.~\ref{sec:background}, we review the Singlet-Doublet model and present expressions for dark sector couplings in the small Yukawa coupling regime. In Sec.~\ref{sec:regimes}, we review different mechanisms of dark matter production, and present the formalism necessary for analysis of these mechanisms.  In Sec.~\ref{sec:results}, we present results and in Sec.~\ref{sec:Signatures} we discuss the experimental signatures. Finally, we conclude.
\section{Singlet-Doublet dark matter}\label{sec:background}

In this work, we study the Singlet-Doublet Majorana fermion model. In this section, we briefly review the structure of the model, the mass spectrum, and couplings.   For reference,  following \cite{Bhattiprolu:2025beq}, we  provide compact expressions for dark sector masses and couplings in the small-Yukawa-coupling limit. 

In the Singlet-Doublet Majorna model, the following left-handed Weyl fermions are added to the SM:  (1) A singlet $S$ in the $({\bf 1}, {\bf 1}, 0)$ representation of the SM $SU(3)_C \times SU(2)_L \times U(1)_Y$ gauge groups, 
and (2) a pair of doublets $D$, and $\overline{D}$  in the $({\bf 1}, {\bf 2}, -\frac{1}{2})$, and $({\bf 1}, {\bf 2}, \frac{1}{2})$ representations  (The bar is part of the field name and does not denote conjugation). To ensure dark matter stability, the new states are odd under a $\mathbb{Z}_2$ symmetry while SM states are even under the same $\mathbb{Z}_2$. This field content is identical to that of the Higgsino and Bino sector of the Minimal Supersymmetric Standard Model (MSSM).  

Following the notation of Ref.~\cite{Bhattiprolu:2025beq}, we identify the $SU(2)_L$ components of the doublets $D$ and $\overline{D}$  as
\begin{equation}
    D = \begin{pmatrix}
        N \\
        C
    \end{pmatrix}
,\qquad
    \overline{D} = \begin{pmatrix}
        \overline{C} \\
        \overline{N}
    \end{pmatrix}
,
\end{equation}
where $N$ and $\overline{N}$ are neutral and $C$ and $\overline{C}$ are charged under $U(1)_{EM}$.

The terms in the Lagrangian that contain the dark sector fields are
\begin{equation}
\begin{aligned}   
    \mathcal{L} = & i S^{\dagger}\overline{\sigma}^{\mu} \partial_{\mu} S + i D^{\dagger}\overline{\sigma}^{\mu}\nabla_{\mu} D + i \overline{D}^{\dagger}\overline{\sigma}^{\mu}\nabla_{\mu}\overline{D} \\
    & - \frac{1}{2} M_S S^2 - M_D D \overline{D}
    - y_1 D H S - y_2 H^{\dagger} \overline{D} S + h.c.,
\end{aligned}
\end{equation}
where  $H \sim ({\bf 1}, {\bf 2}, \frac{1}{2})$ is the SM Higgs doublet, and $M_S, M_D, y_1, y_2$ are the masses and Yukawa couplings of the model, and $\nabla_{\mu}$ is the covariant derivative.  When the Higgs field takes on a non-zero vacuum expectation value, the Yukawa couplings induce mixing between the neutral components of the singlet and doublets. Without loss of generality, we take $M_S, M_D, y_1$ to be positive and real.  We assume CP conservation.  In this case $y_2$ is real. 

The mass matrix for the neutral states $\psi^0 \equiv (S, N, \overline{N})$ resulting from the Lagrangian is 
\begin{equation} \label{eq:mixingmatrix}
     \mathcal{M}_N = 
    \begin{pmatrix}
       M_S & \frac{y_1 v}{\sqrt{2}} & \frac{y_2 v}{\sqrt{2}} \\
       \frac{y_1 v}{\sqrt{2}} & 0 & M_D \\
       \frac{y_2 v}{\sqrt{2}} & M_D & 0
    \end{pmatrix}
    ,
\end{equation}
where $v \simeq 246$ GeV is the  vacuum expectation value of the Higgs field. The dark sector has four mass eigenstates:  three neutral Majorana fermions $\chi^0_i$ ($i =1,2,3$) and one charged Dirac fermion $\chi^{\pm}$.  We take $i = 1$ to denote the lightest neutral state.

\subsection{Couplings and Mass Eigenstates}\label{sec:couplings}

In this work, we focus on the small-Yukawa-coupling regime, i.e., $y_1, |y_2| \ll 1$.  In this limit, the dark matter couplings to the SM are suppressed.  In this section, we present  expressions for masses and couplings of the dark sector eigenstates in this limit.  For discussion of this limit as well as others, see Ref. \cite{Bhattiprolu:2025beq}.   For convenience, we define $y_{+} \equiv y_1 + y_2$ and $y_{-} \equiv y_1 - y_2$.

The masses of the neutral states are approximately given by:
\begin{equation}
\label{eq:SmallY_MassEigenvalues}
    \begin{aligned}
        m_{\chi_a} &\approx M_S - \frac{v^2 y_+^2}{4 \left(M_D - M_S\right)} + \frac{v^2 y_-^2}{4 \left(M_D + M_S\right)},\\
        m_{\chi_b} &\approx M_D + \frac{v^2 y_+^2}{4\left(M_D - M_S\right)},\\
        m_{\chi_c} &\approx M_D + \frac{v^2 y_-^2}{4\left(M_D + M_S\right)}.
    \end{aligned}
\end{equation}
The ordering of $m_{\chi_a},m_{\chi_b},m_{\chi_c}$ depends on the the relative magnitude of $M_{S}$ and $M_{D}$. Explicitly, $a = 1$ when $M_S < M_D$ and $b=1$ when $M_D < M_S$.  In this work, we take $M_{S} < M_{D}$, so  $m_{\chi_1}$ corresponds to the singlet mass, and it is the mass of the dark matter.  The dark sector thus separates into a singlet-like state $\chi^0_{1}$ with approximate mass $M_{S}$  and a trio of doublet-like states, $\chi^0_2, \chi^0_3$ and $\chi^{\pm}$ with approximate mass $M_{D}$.

Next, we turn to couplings relevant for cosmology.
The SM and dark sector couple  via electroweak bosons. 
The coupling structure of the dark sector
to the photon $A_\mu$, $Z$ boson $Z_\mu$,  $W$ bosons $W^\pm_\mu$, and the Higgs boson $h$
is given as \cite{Bhattiprolu:2025beq}:
\begin{equation}
\begin{aligned}   
    \mathcal{L} & \supset
    e A_\mu \overline{\chi^-} \gamma^\mu \chi^- - \frac{e}{s_W c_W} \left(s_W^2 - \frac{1}{2}\right) Z_\mu \overline{\chi^-} \gamma^\mu \chi^-\\
    & - g^V_{W^+ \chi^- \chi^0_i} W^+_\mu \overline{\chi}^0_i \gamma^\mu \chi^-
    - g^A_{W^+ \chi^- \chi^0_i} W^+_\mu \overline{\chi}^0_i \gamma^\mu \gamma_5 \chi^- + c.c.\\
    & - \frac{1}{2} \text{Re}(g_{Z \chi^0_i \chi^0_j}) Z_\mu \overline{\chi}^0_i \gamma^\mu \gamma_5 \chi^0_j
    + \frac{i}{2} \text{Im}(g_{Z \chi^0_i \chi^0_j}) Z_\mu \overline{\chi}^0_i \gamma^\mu \chi^0_j\\
    & - \frac{1}{2} \text{Re}(g_{h \chi^0_i \chi^0_j}) h \overline{\chi}^0_i \chi^0_j
    + \frac{i}{2} \text{Im}(g_{h \chi^0_i \chi^0_j}) h \overline{\chi}^0_i \gamma_5 \chi^0_j,
    \label{eq:SMcouplings}
\end{aligned}
\end{equation}
where
$e$ is the electromagnetic coupling, $s_W (c_W)$ is the (co)sine of the weak mixing angle. We now provide approximate expressions for  relevant couplings enumerated above.  The couplings between the dark sector and the $W$ boson are
\begin{equation} \label{eq:Wcoups_smallY}
\begin{aligned}
g_{W^{+} \chi^{-} \chi_a^0}^V\approx-\frac{g}{4} \frac{vy_+}{\left(M_D-M_S\right)},&\quad  
g_{W^{+} \chi^{-} \chi_a^0}^A\approx-\frac{g}{4} \frac{vy_-}{\left(M_D+M_S\right)},\\
g_{W^{+} \chi^{-} \chi_b^0}^V\approx\frac{g}{2}\left[1-\frac{v^2y_+^2}{8\left(M_D-M_S\right)^2}\right],&\quad
g_{W^{+} \chi^{-} \chi_b^0}^A\approx-\frac{g}{2} \frac{v^2y_+y_-}{8 M_D\left(M_D-M_S\right)},\\
g_{W^{+} \chi^{-} \chi_c^0}^V\approx i \frac{g}{2}\left[1-\frac{v^2y_-^2}{8\left(M_D+M_S\right)^2}\right],&\quad
g_{W^{+} \chi^{-} \chi_c^0}^A\approx-i \frac{g}{2} \frac{v^2y_+y_-}{8 M_D\left(M_D+M_S\right)}.
\end{aligned}
\end{equation}
The couplings of the dark sector particles to the $Z$ boson are
\begin{equation}
\label{eq:Zcoups_smallY}
\begin{aligned}
& g_{Z \chi_a^0 \chi_a^0}\approx\frac{g}{4 c_w} \frac{v^2y_+y_-}{\left(M_D^2-M_S^2\right)}, \\
& g_{Z \chi_b^0 \chi_b^0}\approx-\frac{g}{8 c_w} \frac{v^2y_+y_-}{M_D\left(M_D-M_S\right)}, \\
& g_{Z \chi_c^0 \chi_c^0}\approx -\frac{g}{8 c_w} \frac{v^2y_+y_-}{M_D\left(M_D+M_S\right)}, \\
& g_{Z \chi_a^0 \chi_b^0}\approx -\frac{g}{4 c_w} \frac{vy_-}{\left(M_D+M_S\right)}\approx g_{Z \chi_b^0 \chi_a^0}^*, \\
& g_{Z \chi_b^0 \chi_c^0}\approx i \frac{g}{2 c_w}
\left[1-
\frac{v^2 y_+^2}{8\left(M_D - M_S\right)^2} -
\frac{v^2 y_-^2}{8\left(M_D + M_S\right)^2}
\right]\approx g_{Z \chi_c^0 \chi_b^0}^*, \\
& g_{Z \chi_a^0 \chi_c^0}\approx-i \frac{g}{4 c_w} \frac{vy_+}{\left(M_D-M_S\right)}\approx g_{Z \chi_c^0 \chi_a^0}^*.
\end{aligned}
\end{equation}
The precise form of couplings of the dark sector to the Higgs boson will typically not play a large role in the cosmology but may be found in Ref.~\cite{Bhattiprolu:2025beq}.
   Yukawa couplings ${\gtrsim \mathcal O(10^{-2})}$ are constrained by direct detection, see Sec.~\ref{sec:results}.

Even in the small-Yukawa-coupling limit, some couplings between the gauge bosons and heavier dark sector particles are still full-strength electroweak couplings, i.e. ${\mathcal O} (g)$.  These are the couplings that connect the doublet-like states to one another.  The presence of these full-strength couplings is important.  They enforce chemical equilibrium between the parts of the dark sector that couple in this way, and they also keep this part of the dark sector in kinetic equilibrium. However, in the small-Yukawa-coupling limit, direct detection cross sections are strongly suppressed.

\section{Dark Matter Production Regimes} 
\label{sec:regimes}
Depending on the nature of the interactions between the dark sector particles, as well as between the dark sector and the SM, different dark matter production mechanisms apply.  As we discuss below, in different portions of Singlet-Doublet model parameter space, standard freeze-out (including co-annihilation), co-scattering, freeze-in, and the SuperWIMP mechanism can apply.  Here, we briefly review these mechanisms.

In the \emph{freeze-out} regime, the entire dark sector was in thermal equilibrium with the SM in the early universe.  Then, after the dark matter became non-relativistic, the dark sector eventually decoupled from the SM leaving a cold dark matter relic.  The specific interactions responsible for  decoupling and setting the relic abundance can vary.  One mechanism of particular relevance in the Singlet-Doublet model is co-annihilation \cite{Griest:1990kh,Edsjo:1997bg}.  In the co-annihilation regime, an additional dark sector state close in mass to the dark matter participates in annihilation.  

Our previous work, Ref. \cite{Bhattiprolu:2025beq}, made a detailed study of co-annihilation within the Singlet-Doublet model. In this co-annihilation regime, the dark matter is in chemical equilibrium with states that are doublet-like, and the substantial annihilation cross section of the doublet sector is responsible for the depletion of the dark sector including the dark matter.  Consequently, the dark matter can be very nearly a singlet.   

Another dark matter production mechanism, and one of interest for the present work is \emph{co-scattering/conversion- driven freeze-out} \cite{DAgnolo:2017dbv, Brummer:2019inq, Garny:2017rxs}. The setup is as follows.  Consider two sub-sectors of the dark sector: one composed of $\chi$ the dark matter candidate (for the Singlet-Doublet model $\chi^0_1$) and the other containing $\psi_i$ with $m_{\psi_i} \gtrsim m_{\chi}$ (for the Singlet-Doublet model, $\psi_{i} \in \{ \chi^0_2, \chi^0_3, \chi^{\pm}\}$).  Co-scattering is the process by which conversions between the $\chi$ and $\psi$-sector occur. The $\psi_{i}$ (here, doublets) couple relatively strongly to the SM, and equilbrium between the $\psi$-sector and the SM is easily maintained at early times.  In this co-scattering scenario, $\chi$ is also in equilibrium with the SM in the very early universe by virtue of its interactions with the $\psi$-sector: $\chi X \leftrightarrow \psi X'$ (co-scattering) and/or $\psi_{i} \leftrightarrow \chi X$ (decay).  However, these co-scattering processes can freeze-out before the relic abundance is fully established by other processes, e.g.,  $\psi_{i}\psi_{j} \rightarrow XX^{'}$ (co-annihilation-driven freeze-out). 
If one were to incorrectly assume chemical equilibrium between the dark sub-sectors up until freeze-out, the relic abundance would be underestimated. 

In the Singlet-Doublet model, co-scattering processes are also key to maintaining \emph{kinetic} equilibrium for the dark matter. When kinetic equilibrium is lost, an integrated Boltzmann equation is no longer sufficient to completely characterize the system \cite{Profumo:2025uvx, Binder:2021bmg, Abe:2020obo, Binder:2017rgn, Brummer:2019inq, Garny:2017rxs}. In principle, one must evolve the dark matter distribution function. We will present our methods for solving the full Boltzmann equation in the Singlet-Doublet case in a companion paper \cite{Bhattiprolu:ToAppear}. In that paper, we will demonstrate that a full solution to the Boltzmann equation differs from the integrated one by less than $\lesssim 10\%$ for the relic density, and often much less.\footnote{This conclusion is consistent with generic estimates made in \cite{Garny:2017rxs}.  It also is consistent with estimates \cite{Alguero:2022inz} made for the closely related Singlet-Triplet model, but differs from the calculations performed for that model in \cite{Brummer:2019inq}.} Because this effect is  small, for the remainder of this paper, we will work in the approximation that \emph{kinetic}  equilibrium within the dark sector is maintained.

Another often studied dark matter production mechanism is \emph{freeze-in} \cite{Hall:2009bx}. In the freeze-in scenario, the dark matter $\chi$ is never in chemical equilibrium with the SM.  
The dark matter relic density is slowly built up over time via co-scattering or decay processes.   In the freeze-in regime, decays and co-scattering of heavier dark sector states populate the dark matter until the heavier dark sector states begin to be exponentially suppressed due to Boltzmann-suppression.  At this point, fewer dark sector particles are available to decay (or co-scatter) into dark matter.  In the Singlet-Doublet model, freeze-in can dominate for very small $y_1$ and $y_2$.  Ref. \cite{Calibbi:2018fqf} studied freeze-in within the Singlet-Doublet dark matter model, but with a focus on the case where the singlet was light $\sim {\mathcal O} (10  \rm{\, keV})$.  Such a light dark matter mass ensures the kinematic accessibility of  two-body decays from the doublet sector to the singlet sector, and it also allows the possibility of displaced decays of the doublets  in the LHC detector. In this work, we will primarily focus on weak-scale dark matter, and discuss both the cases where two-body and three-body decays to the dark matter dominate.  Consistent with the findings of \cite{Paul_2026}, we find substantial tension with constraints from Big Bang nucleosynthesis (BBN) \cite{Kawasaki_2018} in the latter case.  

If the decay width of the heavier dark sector states is sufficiently small, the \emph{SuperWIMP} mechanism \cite{Feng:2003xh} may be relevant. In this case,  heavier dark sector particles decay give an appreciable contribution to the dark matter abundance well after they freeze out.  The dark matter relic abundance is inherited from the weak-scale freeze-out of the dark sector particles that ultimately decay. Let $\psi$ denote the heavier dark sector particles and $\chi$ be the dark matter. The SuperWIMP relic density of dark matter is given by $\Omega_{\chi} = \Omega_{\psi}  (m_{\chi}/m_\psi)$, where $\Omega_{\psi}$ is the would-be relic density of the $\psi$ if they were stable.  For both the freeze-in mechanisms and SuperWIMPs, an important constraint comes from requiring that the decays of the heavier states $\psi$ not spoil the successes of BBN.

In the following subsections we present the formalism necessary to describe the evolution of the abundances where these different mechanisms apply.

\subsection{Freeze-Out and Co-Scattering} \label{sec:BoltzmannEq}
The Boltzmann equation describes the evolution of the distribution function of species $i$. In the Friedmann-Robertson-Walker metric it takes the form
\begin{equation}\label{eq:BE_FLRW}
\frac{\partial f_i(p, t)}{\partial t}-H p \frac{\partial f_i(p, t)}{\partial p} = \frac{1}{E}\sum \mathcal{C}[f_i(p, t)]
,
\end{equation}
where $t$ is the cosmic time, $H$ is the Hubble parameter, $p$ is the modulus of momentum, $E$ is the energy, and $\mathcal{C}$ is the collision operator. 
Following the notation of \cite{Alguero:2022inz}, the integrated Boltzmann equations that apply under the assumption of kinetic equilibrium are:

\begin{align}\label{eq:Y1ODE}
\frac{dY_\sect{1}}{dx} &= \frac{1}{3H}\frac{d\hat{s}}{dx}\Bigg[
    \langle \sigma v_{\sect{1}\sect{1}\sect{0}\sect{0}} \rangle (Y_\sect{1}^2-Y_{\sect{1}eq}^2)
    + \langle \sigma v_{\sect{1}\sect{1}\sect{2}\sect{2}} \rangle \left(Y_\sect{1}^2-Y_\sect{2}^2\frac{Y_{\sect{1}eq}^2}{Y_{\sect{2}eq}^2}\right)
    + \langle \sigma v_{\sect{1}\sect{2}\sect{0}\sect{0}} \rangle (Y_\sect{1}Y_\sect{2}-Y_{\sect{1}eq}Y_{\sect{2}eq}) \notag \\
&\quad + \langle \sigma v_{\sect{1}\sect{2}\sect{2}\sect{2}} \rangle 
    \left(Y_\sect{1}Y_\sect{2}-Y_\sect{2}^2\frac{Y_{\sect{1}eq}}{Y_{\sect{2}eq}}\right) 
    - \langle \sigma v_{\sect{1}\sect{2}\sect{1}\sect{1}} \rangle 
    \left(Y_\sect{1}Y_\sect{2}-Y_\sect{1}^2\frac{Y_{\sect{2}eq}}{Y_{\sect{1}eq}}\right) 
    - \frac{\Gamma_{\sect{2}\rightarrow\sect{1}}}{\hat{s}}
    \left(Y_\sect{2}-Y_\sect{1}\frac{Y_{\sect{2}eq}}{Y_{\sect{1}eq}}\right)
\Bigg],
\end{align}

\begin{align}\label{eq:Y2ODE}
\frac{dY_\sect{2}}{dx} &= \frac{1}{3H}\frac{d\hat{s}}{dx}\Bigg[
    \langle \sigma v_{\sect{2}\sect{2}\sect{0}\sect{0}} \rangle (Y_\sect{2}^2-Y_{\sect{2}eq}^2)
    - \langle \sigma v_{\sect{1}\sect{1}\sect{2}\sect{2}} \rangle \left(Y_\sect{1}^2-Y_\sect{2}^2\frac{Y_{\sect{1}eq}^2}{Y_{\sect{2}eq}^2}\right)
    + \langle \sigma v_{\sect{1}\sect{2}\sect{0}\sect{0}} \rangle (Y_\sect{1}Y_\sect{2}-Y_{\sect{1}eq}Y_{\sect{2}eq}) \notag \\
&\quad - \langle \sigma v_{\sect{1}\sect{2}\sect{2}\sect{2}} \rangle 
    \left(Y_\sect{1}Y_\sect{2}-Y_\sect{2}^2\frac{Y_{\sect{1}eq}}{Y_{\sect{2}eq}}\right) 
    + \langle \sigma v_{\sect{1}\sect{2}\sect{1}\sect{1}} \rangle 
    \left(Y_\sect{1}Y_\sect{2}-Y_\sect{1}^2\frac{Y_{\sect{2}eq}}{Y_{\sect{1}eq}}\right) 
    + \frac{\Gamma_{\sect{2}\rightarrow\sect{1}}}{\hat{s}}
    \left(Y_\sect{2}-Y_\sect{1}\frac{Y_{\sect{2}eq}}{Y_{\sect{1}eq}}\right)
\Bigg].
\end{align}
Here, the subscripts $\sect{0}, \sect{1}, \sect{2}$ label the SM sector, the dark matter sector, i.e. $\chi^0_1$, and the remainder of the dark sector $\psi$ (here the doublet sector $\psi_{i} \in \{\chi^0_2,\chi^0_3,\chi^\pm\})$, 
respectively.  The subscript $eq$ denotes the corresponding equilibrium distribution.
$Y_\alpha \equiv n_\alpha/\hat{s} $
are the co-moving number densities 
(for the doublet sector, $n_\sect{2}=n_{\chi^0_2}+n_{\chi^0_3}+n_{\chi^\pm}$), $x\equiv m_{\chi_1}/T$, and $\hat{s}$ is the entropy density.  The thermally averaged cross sections are given by 
\begin{equation}
    \langle \sigma v_{\alpha\beta\gamma\delta}\rangle=\frac{1+\delta_{\alpha\beta}}{n_{\alpha eq}n_{\beta eq}}\sum_{abcd}\frac{1}{1+\delta_{ab}}\frac{Tg_ag_b}{8\pi^4}\int_{s_{min}}^\infty ds\sqrt{s}p_{ab}^2K_1\left( \frac{\sqrt{s}}{T}\right)\sigma(s)_{abcd}
    ,
\end{equation}
where $\delta_{ij}$ is the Kronecker delta symbol, $g_i$ gives the internal spin degrees of freedom, $s$ is the Mandelstam variable, and $\sigma(s)_{abcd}$ is the cross section for the process $ab\rightarrow cd$ averaged over initial states and summed over final states. Greek letters denote the sectors, i.e. $\alpha = \sect{0}, \sect{1}, \sect{2}$, and Latin letters $a\in\alpha$, $b\in\beta$, $c\in\gamma$, $d\in\delta$ denote particles belonging to the sectors. The conversion rate $\Gamma_{\sect{2}\rightarrow\sect{1}}$ includes both decays and co-scattering terms; see e.g. \cite{Alguero:2022inz}:
\begin{equation}\label{eq:reactionrateCE}
    \Gamma_{\sect{2} \rightarrow \sect{1}}=\frac{\sum_{\psi_i \in \sect{2}}\Gamma_{\psi_i\rightarrow\sect{1},\sect{0}} \, g_{\psi_i} m_{\psi_i}^2 K_1(m_{\psi_i}/T)}{\sum_{\psi_i \in \sect{2}}g_{\psi_i} m_{\psi_i}^2K_2(m_{\psi_i}/T)}+\langle \sigma v_{\sect{2}\sect{0}\sect{1}\sect{0}} \rangle n_{\sect{0}eq}
    ,
\end{equation}
where $K_1$ and $K_2$ are the modified Bessel functions of the second kind of orders one and two, respectively. 

When conversion processes are sufficiently rapid, specifically $\Gamma_{\sect{2}\rightarrow\sect{1}}/H\gg1$,  chemical equilibrium is maintained between sectors $\sect{1}$ and $\sect{2}$, and Eqs.~(\ref{eq:Y1ODE}) and (\ref{eq:Y2ODE})  reduce to a single differential equation. So, if this condition holds through the freeze-out epoch, then the total comoving number density $Y=Y_\sect{1}+Y_\sect{2}$ evolves as

\begin{equation}\label{eq:YODE}
    \frac{dY}{dx}=\frac{1}{3H}\frac{d\hat{s}}{dx}\langle \sigma v_{eff} \rangle(Y^2-Y_{eq}^2),
    \qquad\textrm{(Co-annihilation)}
\end{equation}
with the effective cross section  $\langle \sigma v_{eff} \rangle $ given by \cite{Griest:1990kh,Edsjo:1997bg}
\begin{equation}\label{eq:sigmaVeff}
    \langle \sigma v_{eff} \rangle =\langle \sigma v_{\sect{1}\sect{1}\sect{0}\sect{0}}\rangle\frac{Y_{\sect{1}eq}^2}{Y_{eq}^2}+\langle  \sigma v_{\sect{1}\sect{2}\sect{0}\sect{0}}\rangle\frac{Y_{\sect{1}eq}Y_{\sect{2}eq}}{Y_{eq}^2}+\langle \sigma v_{\sect{2}\sect{2}\sect{0}\sect{0}}\rangle\frac{Y_{\sect{2}eq}^2}{Y_{eq}^2}.
\end{equation}
Our interest will be in cases where only the last term in Eq.~(\ref{eq:sigmaVeff}) makes a significant contribution (this  minimizes constraints from direct detection).  When $Y_\sect{1}$ and $Y_\sect{2}$  must be tracked separately, but again annihilations involving $\chi^0_1$ are negligible, Eqs.~(\ref{eq:Y1ODE}) and (\ref{eq:Y2ODE}) simplify to: 
\begin{equation} \label{eq:Y1simpleODE}
\left.
\begin{aligned}
\frac{dY_\sect{1}}{dx} 
&= -\frac{1}{3H}\frac{d\hat{s}}{dx}
    \frac{\Gamma_{\sect{2}\rightarrow\sect{1}}}{\hat{s}}
    \Bigg[Y_\sect{2} - Y_\sect{1}\frac{Y_{\sect{2}eq}}{Y_{\sect{1}eq}}\Bigg],  \qquad  \\[6pt]
\frac{dY_\sect{2}}{dx} 
&= \frac{1}{3H}\frac{d\hat{s}}{dx}\Bigg[
    \langle \sigma v_{\sect{2}\sect{2}\sect{0}\sect{0}} \rangle (Y_\sect{2}^2 - Y_{\sect{2}eq}^2)
    + \frac{\Gamma_{\sect{2}\rightarrow\sect{1}}}{\hat{s}}
    \left(Y_\sect{2} - Y_\sect{1}\frac{Y_{\sect{2}eq}}{Y_{\sect{1}eq}}\right)
\Bigg].
\end{aligned}
\right\} 
\textrm{(Co-scattering)}
\end{equation}
Omitting direct $\chi^0_1$ annihilation in this way is an excellent approximation in the parameter space of interest for the Singlet-Doublet model.

\subsection{Freeze-In}
If sufficiently weakly coupled, the dark matter never reaches thermal equilibrium with the SM bath, though the rest of the dark sector may. 
The dark matter abundance is gradually built up in the early universe via a variety of processes.  This can include decays of those dark sector particles that are in equilibrium into the dark matter, the annihilation of SM particles into dark matter, and co-scattering processes where dark sector particles are converted to dark matter.  While dark sector decays into the dark matter are often most relevant during the epoch where the dark sector states take on their full thermal abundance, long-lived dark sector  states can contribute a so-called ``SuperWIMP" contribution when the late decays of the``frozen-out" dark sector abundance contributes significantly to the dark matter abundance. 

Our interest is in the case where the freeze-in abundance is determined by decays and scatterings to the dark matter from a dark sub-sector that is in chemical equilibrium with the SM.   To account for these processes one solves the system Eqs.~(\ref{eq:Y1simpleODE}) 
with initial conditions $Y_\sect{1}=0$ and $Y_\sect{2}=Y_{\sect{2}eq}$ imposed when $T\gg M_D$. An  approximate solution may be obtained by fixing $Y_\sect{2}=Y_{\sect{2}eq}$ and neglecting $Y_\sect{1}$ on the right-hand side of the first line of Eq.~\eqref{eq:Y1simpleODE}. In this approximation, the freeze-in abundance is given by
\begin{equation} \label{eq:freezeinevolution}
    Y_\sect{1}=-\int_0^\infty dx \,\frac{1}{3H}\frac{d\hat{s}}{dx}\frac{\Gamma_{\sect{2}\rightarrow\sect{1}}}{\hat{s}}Y_{\sect{2}eq}
    .
    \qquad \textrm{(Freeze-in)}
\end{equation}
Note, $\Gamma_{\sect{2} \rightarrow \sect{1}}$ contains contributions from both decays and scattering processes; see Eq.~(\ref{eq:reactionrateCE}). Unless decays are suppressed, as is the case if only three-body decays are allowed at tree-level, then the contribution from decays typically dominates.

This decay contribution to the freeze-in relic abundance can be compactly written as \cite{Hall:2009bx}
\begin{equation}\label{eq:FreezeInAbundance}
\Omega_{FI}h^2=m_{\chi_1}\frac{s_0h^2}{\rho_c}\frac{135M_{Pl}}{8\pi^3(1.66)g^{3/2}_{*}}\sum_{{\psi_i}\in\sect{2}}\frac{g_{\psi_i}\Gamma_{{\psi_i}\rightarrow\sect{1},\sect{0}}}{m_{\psi_i}^2},
\end{equation}
where $m_{\chi_1}$ is the dark matter mass, $s_0$ is the present day entropy density, $\rho_c$ is the critical density, and $M_{\rm Pl}$ is the Planck mass. The sum runs over dark sector particles that decay into dark matter. Here, $\Gamma_{\psi_i \rightarrow \sect{1},\sect{0}}$ denotes the partial decay width of $\psi_i$ into $\chi^0_1$, $g_{\psi_i}$ are the internal degrees of freedom, and $m_{\psi_i}$ is the mass of the decaying dark sector particle. In writing Eq.~(\ref{eq:FreezeInAbundance}), we have taken the number of entropic and energy-density relativistic degrees of freedom to be equal and denoted their common value by $g_*$, evaluated at the characteristic production temperature $T \sim m_{\psi_i}$.  
The contribution to the dark matter abundance from the SuperWIMP mechanism may be written as \cite{Feng:2003xh}: 
\begin{equation} \label{eq:SWAbundance}
\Omega_{SW}h^2=\sum_{\psi_i\in\sect{2}}\Omega_{\psi_i}h^2\frac{m_{\chi_1}}{m_{\psi_i}},
\end{equation}
where again the sum runs over particles that decay to dark matter. Here $\Omega_{\psi_i}h^2$ is the would-be relic density of $\psi_i$ if they were stable. In the Singlet-Doublet model considered here, $m_{\chi_1}\simeq M_S$, while the sum runs over the predominantly doublet-like states $\chi^0_2$, $\chi^0_3$, and $\chi^\pm$, whose masses are approximately degenerate, $m_{\psi_i}\simeq M_D$.

In the case where co-scattering in the early universe makes a significant contribution to freeze-in, care  should be taken to properly handle the $t$-channel singularity. \cite{Alguero:2023zol,Grzadkowski_2022,Melnikov_1997}. 
This singularity is regulated by a thermal contribution to the mediator's mass
\begin{equation}
m_v^2(T)=m_v^2+\kappa_v^2T^2,
\end{equation}
which arises when the mediator is in thermal equilibrium with the bath. The relevant values of $\kappa$ are given in Ref.~\cite{Pearce_2022} in the limit of $T\rightarrow \infty$:  
\begin{eqnarray}
\kappa_Z^2 &=& \kappa_W^2
= \frac{11}{6}g_2^2
\approx 0.779, \nonumber \\
\kappa_\gamma^2
&=& \frac{11}{6}g_1^2
\approx 0.234, \\
\kappa_h^2
&=& \frac{3g_2^2+g_1^2}{16}
+ \frac{\lambda}{2}
+ \frac{y_t^2}{4}
\approx 0.372. \nonumber
\end{eqnarray}
If this thermal mass is not properly included, a computation of freeze-in can give the false impression that freeze-in is UV-dominated and dependent on the reheat temperature $T_{RH}$. Properly including the thermal mass for the mediators ensures the dominant co-scattering freeze-in contribution occurs at $T\sim m_{\psi_i}$.  The abundance from co-scattering is thus independent of the precise reheat temperature  $T_{RH}$, as long as $T_{RH}$ exceeds  $m_{\psi_i}$. 

In summary, the freeze-in dark matter relic abundance receives several contributions. Dark matter can be produced via the decay of heavier dark-sector particles to the dark matter, SM annihilations to the dark matter, and co-scattering processes. After the heavier dark sector states freeze out,  their late decays provide an additional SuperWIMP contribution to the dark matter relic density which may or may not be significant. While these contributions are in principle included via numerical solution of the coupled Boltzmann equations,  
Eqs.~\eqref{eq:FreezeInAbundance} and~\eqref{eq:SWAbundance} provide useful analytic estimates of the early decay and SuperWIMP components, respectively. We will discuss below in detail when we expect these different estimates to be relevant to the Singlet-Doublet model.

\section{Results from Cosmology} 
\label{sec:results}

In this section, we move away from generalities and focus on the Singlet-Doublet model. We characterize the parameter space of the model that gives rise to the correct relic density, and we discuss the details of how that relic density comes about in both freeze-out and freeze-in regimes.  

Our focus is on dark matter mass candidates in the 100 GeV to $\sim$ TeV range. 
 We consider the CP-conserving case and take $M_S$, $M_D$, and $y_1$ to be real and positive, while $y_2$ is allowed to take either sign. We restrict the ratio $y_2/y_1$ to the interval $[-1,1]$, which, after accounting for the interchange symmetry between the two Yukawa couplings, spans the physically distinct possibilities.
 As a reminder, our focus is on the small Yukawa coupling regime, $y_1, |y_2| \ll 1$ and $M_{S} < M_{D}$. The latter condition means that dark matter is predominantly singlet-like and $m_{\chi_{1}} \simeq M_{S}$.  When Yukawa couplings are small,   direct detection constraints are not significant.
 The evolution of the abundance of the singlet-like dark matter depends primarily on its interactions with the predominantly doublet-like states $\chi^0_2$, $\chi^0_3$, and $\chi^\pm$. 

Our analysis utilized \texttt{SARAH v4.15.2} \cite{Staub:2008uz, Staub:2013tta, Staub:2015kfa,Vicente:2015zba}, \texttt{SPheno v4.0.5} \cite{Porod:2011nf,Porod:2003um}, and \texttt{micrOMEGAs v6.0}\cite{Alguero:2023zol}. For computations of the relic density, tree-level expressions for the masses are sufficient.  Loop effects are relevant for determining the mass splittings between the nearly degenerate doublet states with mass $\sim M_D$ \cite{Thomas:1998wy}. These splittings play an important role in the collider phenomenology discussed in Sec.~\ref{sec:Signatures}.

Dark matter production is primarily sensitive to the magnitude of the Yukawa couplings and the mass difference $\Delta m \equiv M_D-M_S$.  When Yukawa couplings are sufficiently large, chemical equilibrium between the dark matter and the doublet sector is maintained through the epoch of co-annihilation freeze-out.  In this case, the relic abundance $\Omega h^2$ is determined by the standard co-annihilation mechanism. For smaller Yukawa couplings, chemical equilibrium between the dark matter and doublet sector is lost before co-annihilation freeze-out. In this regime, co-scattering processes must be consistently included to correctly calculate the relic abundance. Finally, for sufficiently small Yukawa couplings, the dark matter never reaches chemical equilibrium with the doublet sector (or the SM bath). Then, dark matter may be generated through freeze-in, with contributions from doublet decays, SM annihilation, and co-scattering processes.

Our results in this section are divided into two parts. We first consider the parameter space in which the dark matter relic abundance is established through freeze-out, including both co-annihilation and co-scattering effects. We then turn to the freeze-in regime, in which the dark matter abundance is generated through feeble interactions with the thermal bath. 

\subsection{Freeze-out regime}
\label{sec:FOregime}

We now examine the different mechanisms that determine the dark matter relic abundance when the dark matter reaches chemical equilibrium with the SM in the early Universe.   Depending on the precise values of the Yukawa couplings and the mass splitting $\Delta m$, chemical equilibrium may either
\begin{itemize}
\item[] {(i) persist until the relic density is  established via co-annihilation driven freeze-out or} 
\item[]{(ii) break down while the relic abundance is still being established. }
\end{itemize}

\begin{figure}[h!]    \includegraphics[width=\linewidth]{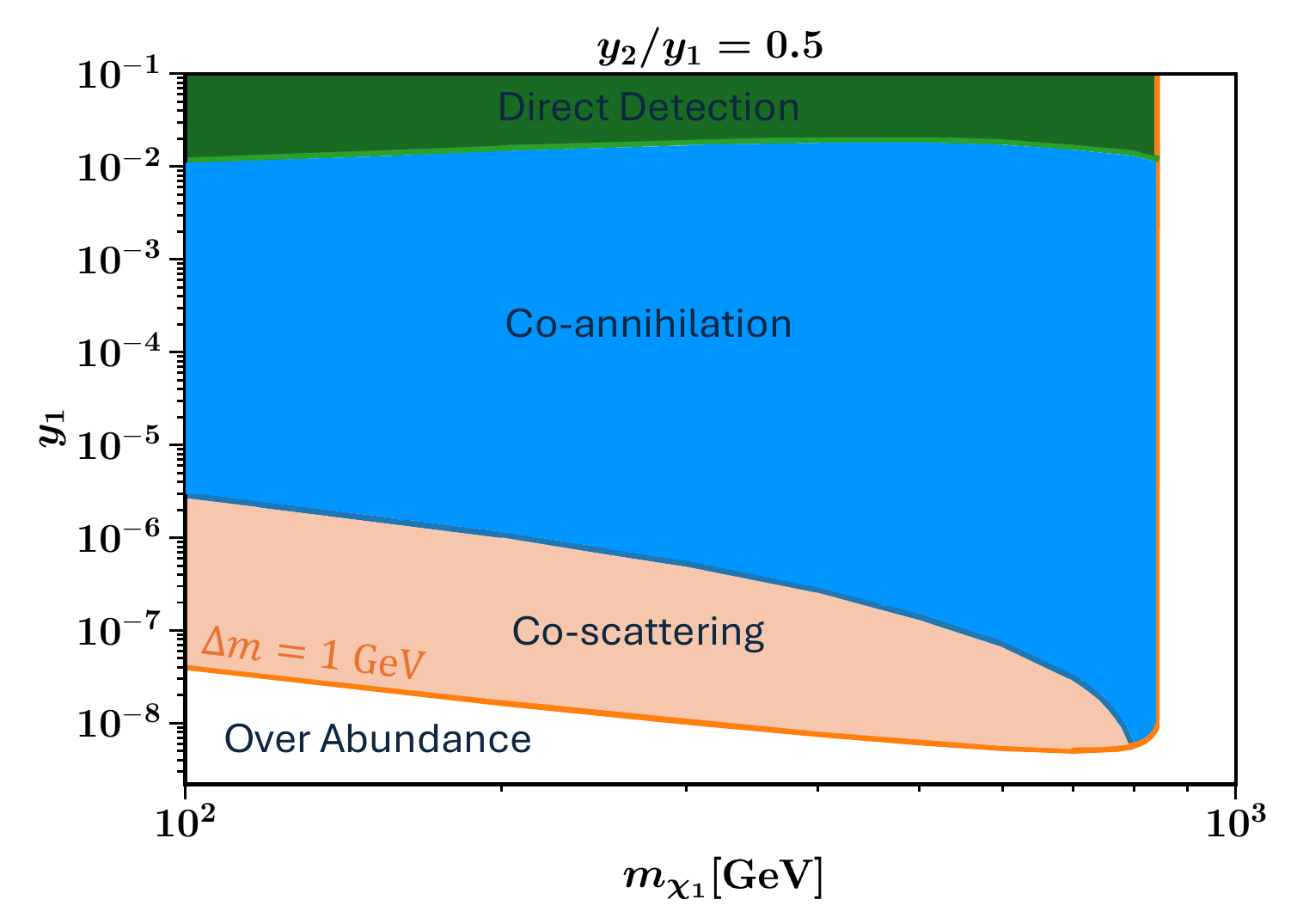}
    \caption{Parameter space for freeze-out dark matter production. All points in the shaded region can realize the correct relic abundance for some choice of the mass splitting $\Delta m$ between $M_{S}$ and $M_{D}$.  In the top region (green) there is a too-large direct detection cross-section mediated via Higgs boson exchange. The blue region corresponds to the co-annhilation regime studied, e.g., in Ref.~\cite{Bhattiprolu:2025beq}. Below this, chemical equilibrium between the dark matter and the doublets is lost prior to freeze-out (orange).   Below the orange region, freeze-out happens too early, and dark matter over-closes the universe.  The location of this boundary depends on the choice of the mass splitting $\Delta m$, here 1 GeV.  For much lower Yukawa couplings, there is a potential possibility of a freeze-in history, see Sec.~\ref{sec:FIRegime}}
    \label{fig:parameterSpace}
\end{figure}

In Fig.~\ref{fig:parameterSpace} we show which dark matter mechanisms operate in the $m_{\chi_1}$ vs. $y_{1}$ plane for a fixed ratio of Yukawa couplings $y_{2}/y_{1}=0.5$.  One purpose of this section will be to understand this plot and these mechanisms in some detail.  For now, we discuss the broad features.  At the largest couplings  $y_1 \gtrsim 10^{-2}$ (green region), spin-independent direct detection signals, mediated by Higgs boson exchange, exclude the model.  The blue region  (approximately to $10^{-6}\lesssim y_1\lesssim 10^{-2}$) corresponds to cosmology of the type (i), where chemical equilibrium with the doublet sector is well-maintained throughout the evolution prior to freeze-out.  In the orange region this equilibrium breaks down; cosmology of type (ii) applies, and understanding the details of co-scattering becomes relevant.   At sufficiently small couplings (below the boundary of the orange region), the dark matter would freeze-out very early and would be over-abundant.  At substantially smaller couplings than those shown in this figure, $y_1 \lesssim 10^{-12}$, it is possible that the dark matter was never in equilibrium.  This is the subject of the freeze-in discussion in the next section.

We have defined the boundary between the co-annihilation (blue) and co-scattering regions (orange) as the point at which the relic-density predictions obtained using the single ODE of Eq.~\eqref{eq:YODE} and the two ODE system of Eq.~\eqref{eq:Y1ODE} and \eqref{eq:Y2ODE} begin to differ by at least 10\%.  This discrepancy signals the breakdown of chemical equilibrium between the dark matter and the doublet sector. Within the orange region, the coupled ODE system is therefore required to correctly determine the relic abundance.  Moving below the boundary, the difference between the two treatments can be much larger, and an incorrect application of the single ODE can cause errors in the computation of $\Omega h^2$ by more than 100\%.

The boundary between the co-scattering and overabundant region is shown for a minimum mass splitting of $\Delta m = 1~\mathrm{GeV}$. A larger (smaller) mass-splitting would shift the boundary to larger (smaller) Yukawa couplings. Generally, the dominant coupling that contributes to scattering (and hence determines the boundary) is proportional to $vy_+/\Delta m$, where $y_+=y_1+y_2$. So, decreasing $\Delta m$ corresponds to smaller values of Yukawa couplings for a fixed co-scattering rate. However, there is no reason that $M_{S}$ and $M_{D}$ should be very close to one another; indeed, even if they start at similar values in the UV, their different quantum numbers tend to push these masses apart under renormalization group evolution, see, e.g., \cite{Bhattiprolu:2025beq}.

The cutoff near $m_{\chi_1}\simeq 850~\mathrm{GeV}$ marks the transition from predominantly singlet-like to predominantly doublet-like dark matter. For predominantly singlet-like dark matter with $m_{\chi_1}\gtrsim 850~\mathrm{GeV}$, the freeze-out mechanism considered here always overproduces the dark matter relic abundance.  Above this mass, doublet-like dark matter is, however, possible.  The phenomenology mimics that of pure Higgsino dark matter.

\begin{figure}[t]
    \centering

    \begin{subfigure}{0.48\textwidth}
        \centering
        \includegraphics[width=\linewidth]{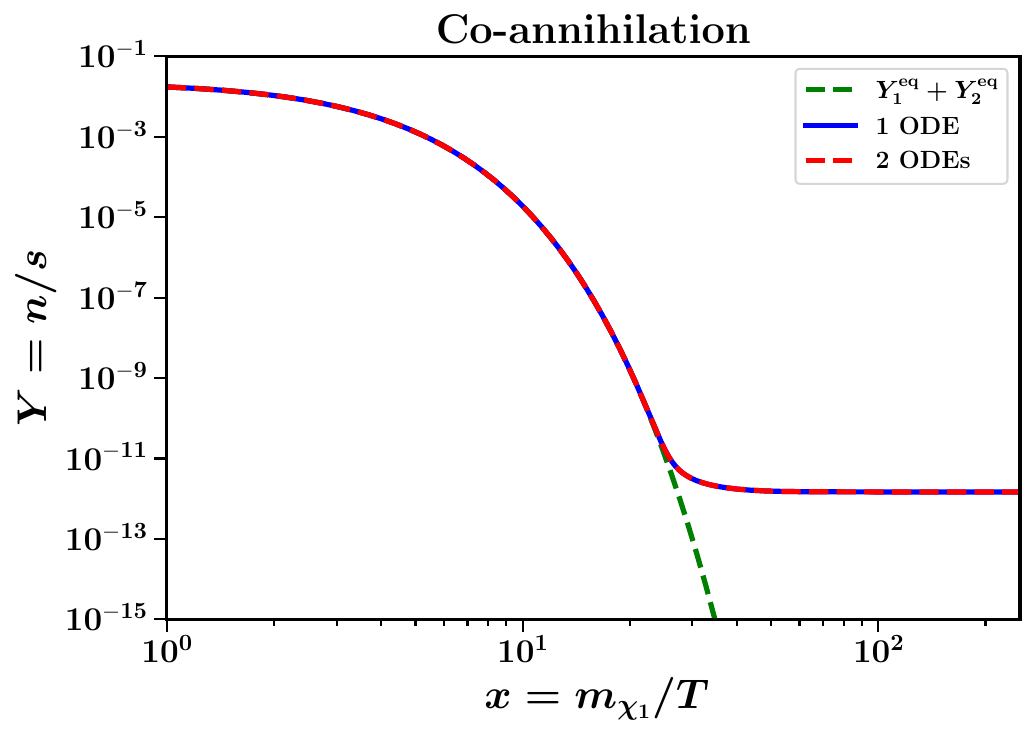}
    \end{subfigure}
    \hfill
    \begin{subfigure}{0.48\textwidth}
        \centering
        \includegraphics[width=\linewidth]{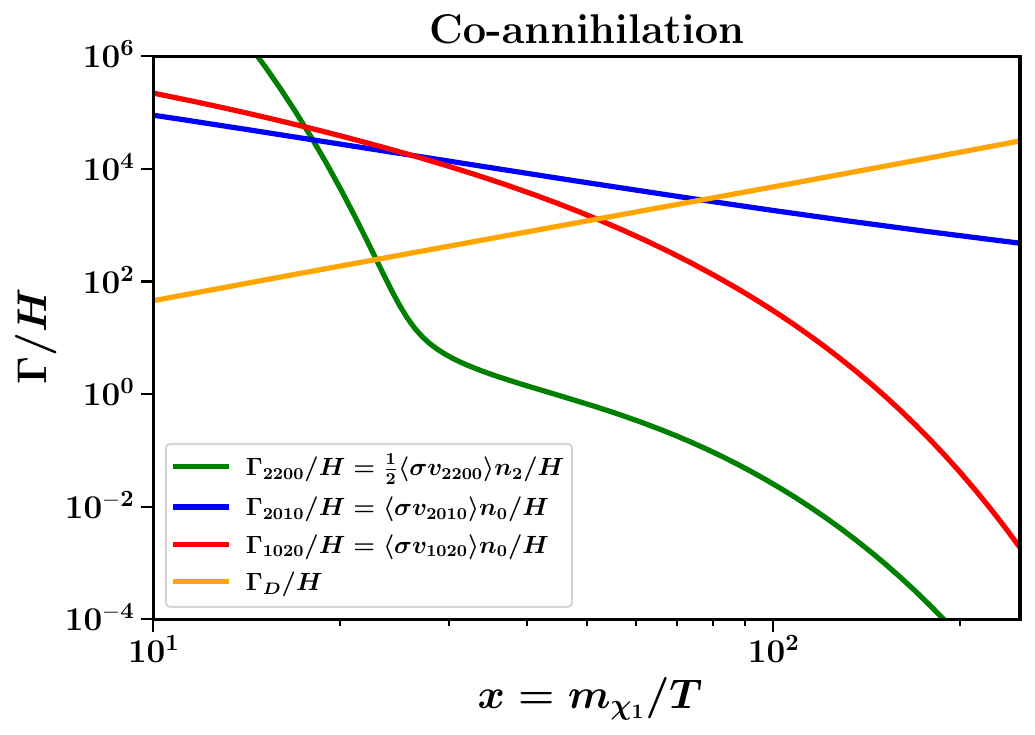}
    \end{subfigure}

    \begin{subfigure}{0.48\textwidth}
        \centering
        \includegraphics[width=\linewidth]{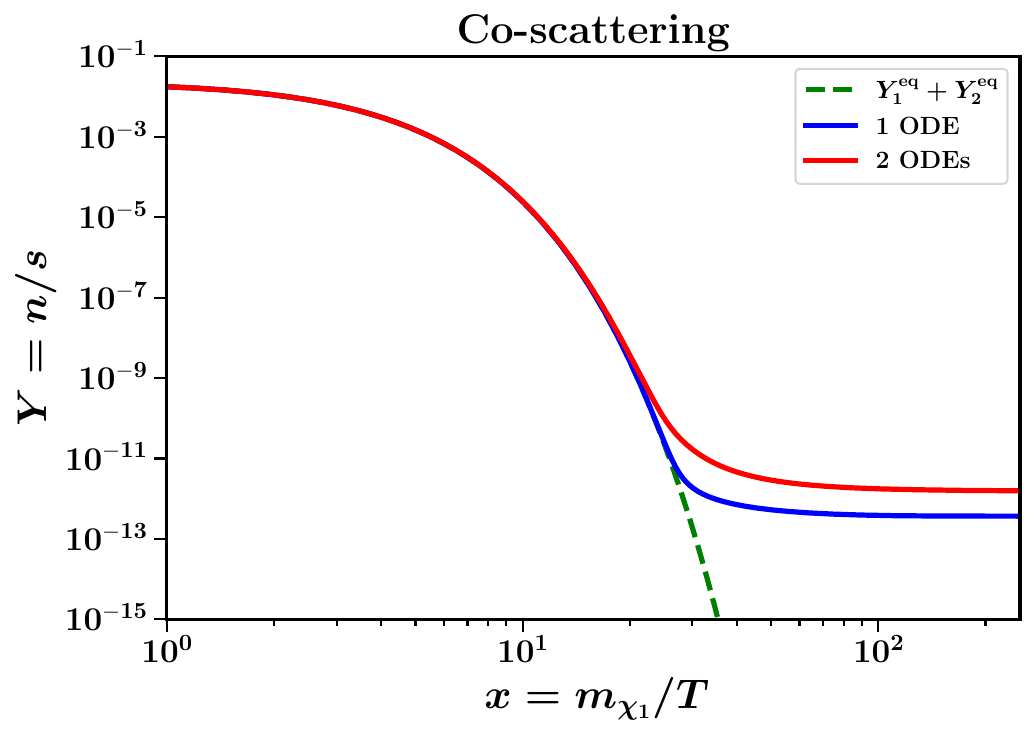}
    \end{subfigure}
    \hfill
    \begin{subfigure}{0.48\textwidth}
        \centering
        \includegraphics[width=\linewidth]{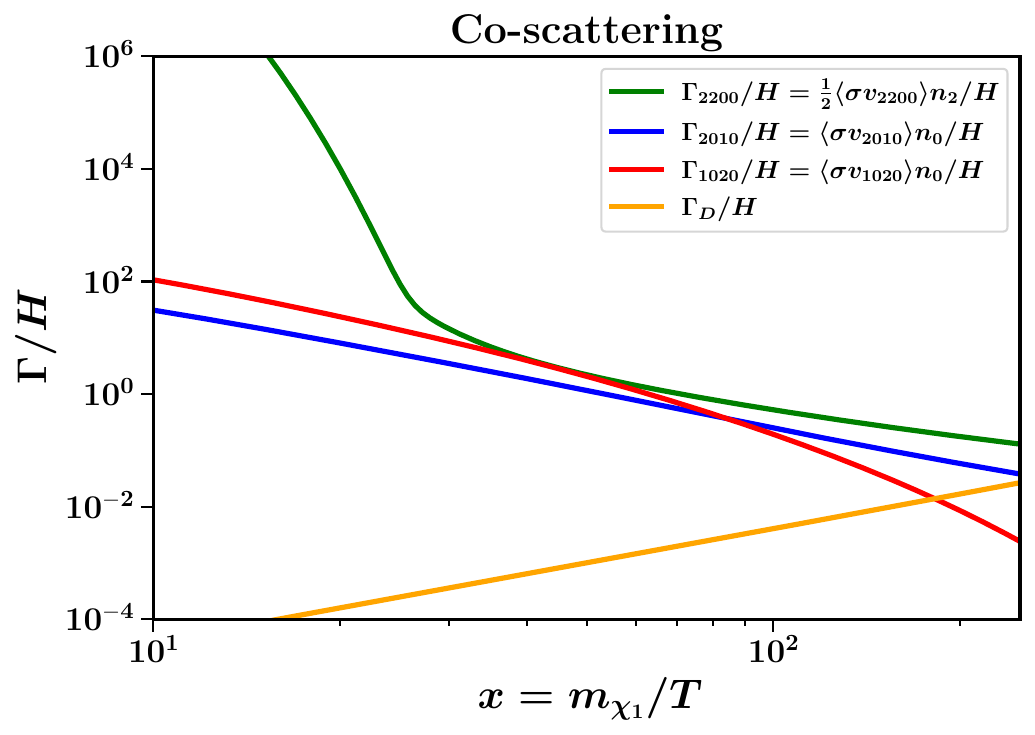}
    \end{subfigure}

    \caption{Left: Evolution of the comoving abundance, $Y=Y_\sect{1}+Y_\sect{2}$ as a function of $x \equiv m_{\chi_{1}}/T$ during freeze-out production.  The evolution is shown for both a relatively large value of the Yukawa coupling $y_{1}=10^{-5}$ for which chemical equilibrium obtains throughout (top panels, labeled co-annihilation) and also for a smaller value of the Yukawa coupling $y_{1} = 6 \times 10^{-8}$ (bottom panels, labeled co-scattering), where the assumption of chemical equilibrium between the singlet and doublet sector breaks down prior to freeze-out. 
    Right: Reaction rates normalized to the Hubble expansion rate $\Gamma/H$ as a function of $x$ for $y_{1}=10^{-5}$ (top) and $y_{1} = 6 \times 10^{-8}$ (bottom).   Shown are co-annihilation rates of the doublet sector to the SM $\Gamma_{\sect{2}\sect{2}\sect{0}\sect{0}}$ (green), co-scattering  between the doublet and singlet sector $\Gamma_{\sect{2}\sect{0}\sect{1}\sect{0}}$ (blue) and  $\Gamma_{\sect{1}\sect{0}\sect{2}\sect{0}}$ (red) and decays $\Gamma_{D} \equiv \Gamma_{\chi_2}+ \Gamma_{\chi_3}+\Gamma_{\chi^{\pm}}$ from the doublet-sector to singlet sector (orange). For all four of these plots, $M_S$ was fixed at 300 GeV and $y_2/y_1=0.5$.
    }
    
    \label{fig:YvsXFO}
\end{figure}

We next illuminate the physics underlying the cosmology of types (i) and (ii) using representative benchmark points. The two left panels of Fig.~\ref{fig:YvsXFO} show the evolution of the total comoving abundance for the sum of the dark matter and the doublet sector, i.e. $Y=Y_\sect{1}+Y_\sect{2}$, for two choices of $y_{1}$, fixing $M_{S}=300~\mathrm{GeV}$ and $y_2/y_1=0.5$. For each value of $y_1$, the doublet mass $M_D$ is chosen such that the solution of the coupled Boltzmann equations reproduces the observed relic abundance, $\Omega h^2=0.12$. The blue curves show the solution of the single effective Boltzmann equation, Eq.~(\ref{eq:YODE}), referred to as the ``1-ODE treatment", which assumes chemical equilibrium between the dark matter and the doublet states. The red curves show the total abundance $Y_\sect{1}+Y_\sect{2}$ obtained by solving the coupled equations, Eqs.~(\ref{eq:Y1ODE}) and (\ref{eq:Y2ODE}), referred to as the ``2-ODE treatment". The dashed green curves denote the total equilibrium abundance, $Y_\sect{1}^{\rm eq}+Y_\sect{2}^{\rm eq}$.

For $y_1=10^{-5}$ (upper panels), chemical equilibrium throughout the full dark sector is maintained up through the epoch of freeze-out (type (i)), and co-annihilation determines the relic abundance. For such a history, the 1-ODE and 2-ODEs treatments give identical evolutions for the abundances. For $y_1=6\times10^{-8}$ (lower panels), however, the dark matter and doublet states fall out of chemical equilibrium near the freeze-out epoch (type (ii)). The two treatments diverge, and using a single effective equation underestimates the final abundance by approximately a factor of four. 

The processes responsible for maintaining chemical equilibrium between two sectors include co-scattering and decay processes. Qualitatively, chemical equilibrium is lost when the reaction rates of these processes given in Eq.~(\ref{eq:reactionrateCE}) are slow compared to the Hubble expansion rate. Consequently, the 1-ODE treatment suffices when $\Gamma_{\sect{2}\rightarrow\sect{1}}/H\gg 1$ throughout the entire period preceding freeze-out via co-annihilation; the freeze-out condition is   approximately determined by $\Gamma_{\sect{2}\sect{2}\sect{0}\sect{0}}/H\sim \mathcal{O}(1)$, which occurs at $x\sim25$. 

The two right panels of Fig.~\ref{fig:YvsXFO} show reaction rates normalized by the Hubble expansion rate $H$. The co-annihilation rate $\Gamma_{\sect{2}\sect{2}\sect{0}\sect{0}}=\frac{1}{2}\langle \sigma v_{\sect{2}\sect{2}\sect{0}\sect{0}}\rangle n_\sect{2}$ is proportional to the number density of the doublets $n_\sect{2}$; see  \cite{P_Gondolo_2004} for a discussion of the factor of $1/2$. The co-scattering rate responsible for converting doublets to dark matter and the reverse process are given by $\Gamma_{\sect{2}\sect{0}\sect{1}\sect{0}}=\langle \sigma v_{\sect{2}\sect{0}\sect{1}\sect{0}}\rangle n_\sect{0}$ and $\Gamma_{\sect{1}\sect{0}\sect{2}\sect{0}}=\langle \sigma v_{\sect{1}\sect{0}\sect{2}\sect{0}}\rangle n_\sect{0}$, respectively. Here $n_\sect{0}$ is the number density of SM particles and the co-scattering rates are related by $\Gamma_{\sect{1}\sect{0}\sect{2}\sect{0}}=\frac{n_{\sect{2}eq}}{n_{\sect{1}eq}}\Gamma_{\sect{2}\sect{0}\sect{1}\sect{0}}$. Here $n_{\sect{1}eq}$ and $n_{\sect{2}eq}$ are the equilibrium number density for sectors $\sect{1}$ and $\sect{2}$ respectively. $\Gamma_{D}=\Gamma_{\chi_2}+\Gamma_{\chi_3}+\Gamma_{\chi^{\pm}}$ is the total decay width of the doublet states to dark matter.
For all points in the freeze-out regime (including the ones shown here), decay processes are subdominant compared to the co-scattering process at the time of freeze-out. 
For $y_1=10^{-5}$ (top), we see that co-scattering and decay processes are sufficiently rapid during the freeze-out epoch, $x\sim25$, to ensure chemical equilibrium after co-annihilation-driven freeze-out. In contrast, for $y_1=6\times10^{-8}$, the ratio of the conversion rate to  the Hubble expansion rate is close to unity near freeze-out. This indicates the necessity of the 2-ODE treatment.

\begin{figure}[t]
    \centering
    \begin{subfigure}{0.48\textwidth}
        \centering
        \includegraphics[width=\linewidth]{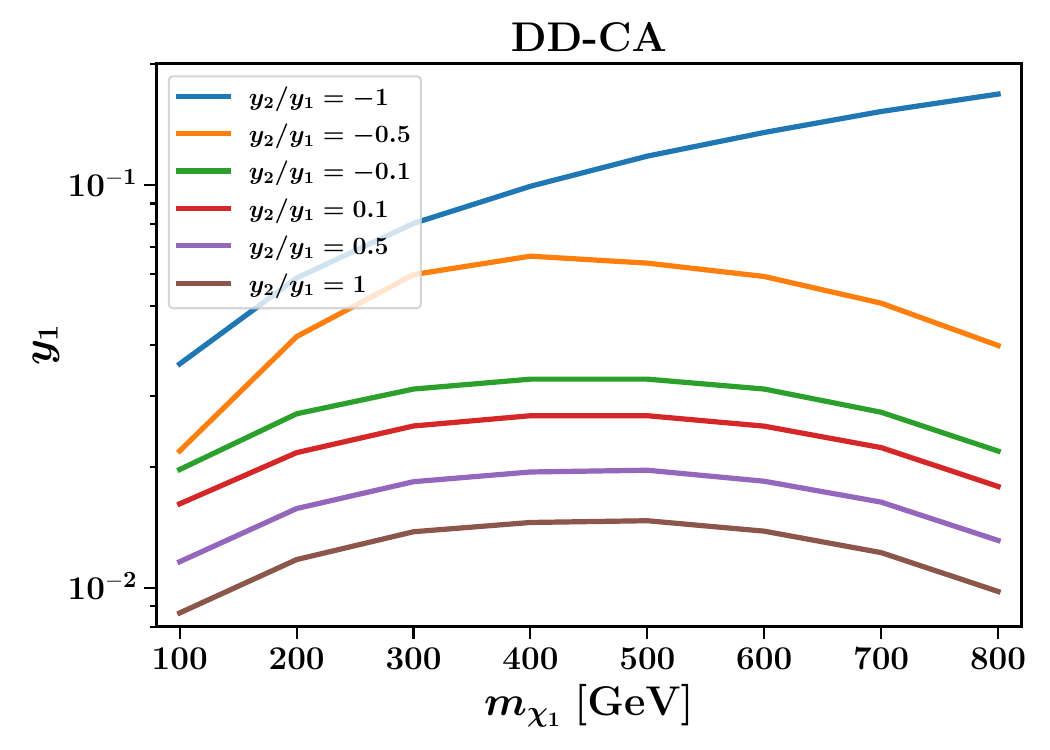}
    \end{subfigure}
    \hfill
    \begin{subfigure}{0.48\textwidth}
        \centering
        \includegraphics[width=\linewidth]{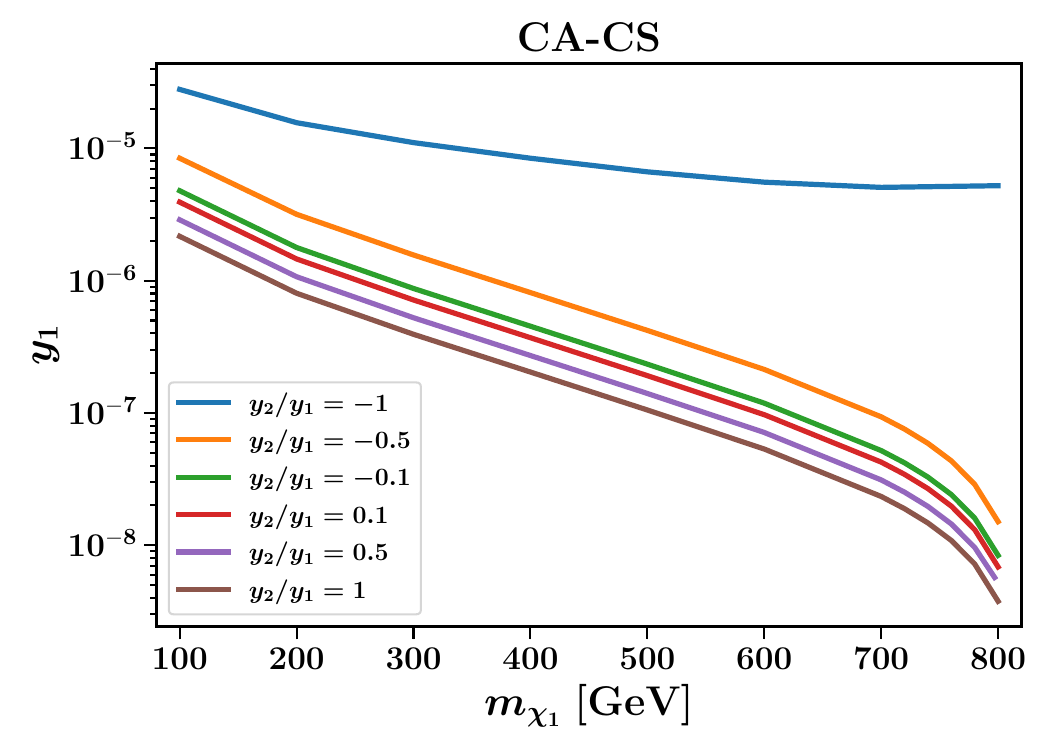}
    \end{subfigure}

    \vspace{0.35cm}

    \begin{subfigure}{0.48\textwidth}
        \centering
        \includegraphics[width=\linewidth]{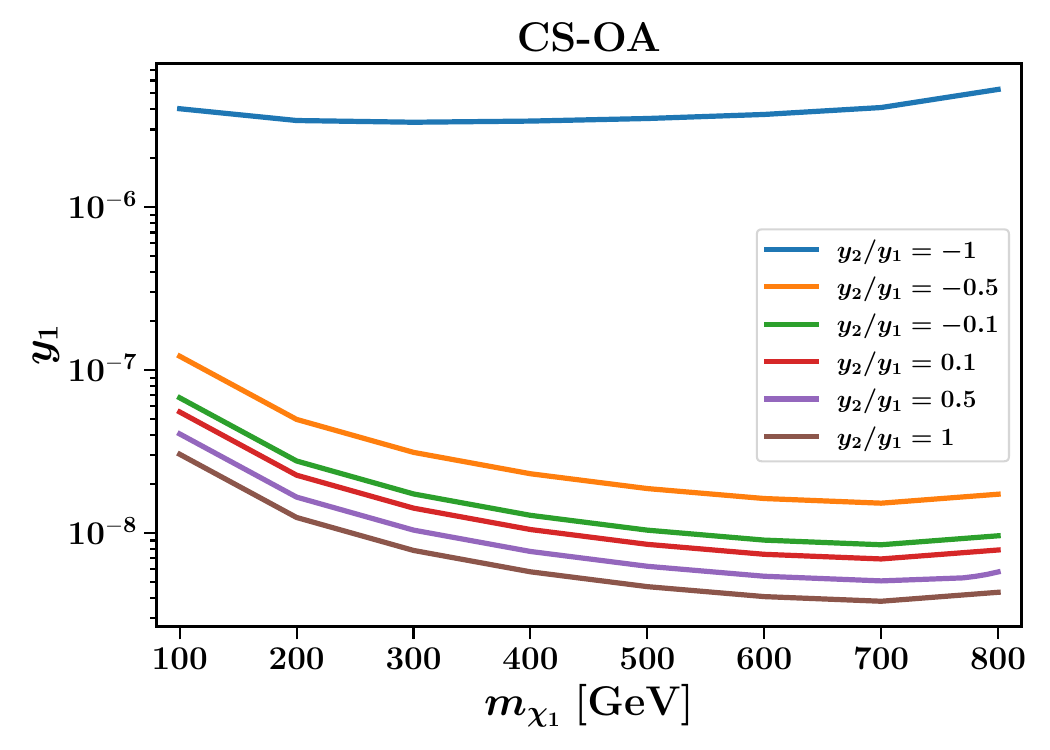}
    \end{subfigure}

    \caption{Dependence of the boundaries between the different dark matter production regimes on the Yukawa ratio $y_2/y_1$. Shown are the direct-detection/co-annihilation boundary (DD--CA, upper left), the co-annihilation/co-scattering boundary (CA--CS, upper right), and the co-scattering/overabundance boundary (CS--OA, lower panel).}
    \label{fig:ShapeOfParameterSpace}
\end{figure}

We next explore how variation in the ratio of ratio the Yukawa couplings $y_{2}/y_{1}$ impacts the shapes of the boundaries between the dark matter generation regimes shown in Fig.~\ref{fig:parameterSpace}.  There, the boundaries were shown for a fixed $y_{2}/y_{1}= 0.5$. 
The effect of varying this ratio is shown in Fig.~\ref{fig:ShapeOfParameterSpace}. The boundaries between the direct-detection and co-annihilation regions, the co-annihilation and co-scattering regions, and the co-scattering and overabundant regions are denoted by DD--CA, CA--CS, and CS--OA, respectively.
Though the value of $y_{2}/y_{1}$ affects where a given boundary appears in $y_{1}$ space, the choice of $y_2/y_1$  has little impact on the shape of the boundaries. except near $y_2/y_1\sim -1$. At this special point, which conserves a custodial symmetry, $y_{+} \rightarrow 0$, while $y_{-}$ remains of order $y_{1}$.  Importantly, the dark sector possesses some couplings that scale as $y_{+}/\Delta m$  and others that scale $y_{-}/(M_S+M_D)$, see Sec. ~\ref{sec:couplings}, Eqs.(\ref{eq:Wcoups_smallY}), (\ref{eq:Zcoups_smallY}).   Away from $y_{2}/y_{1} \rightarrow -1$,  the $y_{+}$ couplings dominate the dark matter detection and generation processes.  But for Yukawa coupling ratios very close to $-1$, the couplings proportional to $y_{-}$ dominate (in spite of the fact that they are parametrically suppressed by the factor $\Delta {m}/(M_S+M_D)$). Had we shown plots showing the $y_{+}$ value at the boundary rather than the $y_{1}$ value, the location of the $y_{+}$ boundaries for different $y_2/y_1 \neq 1$ would be nearly identical.

\begin{figure}[t]
    \centering

    \begin{subfigure}[t]{0.48\textwidth}
        \centering
        \includegraphics[width=\linewidth]{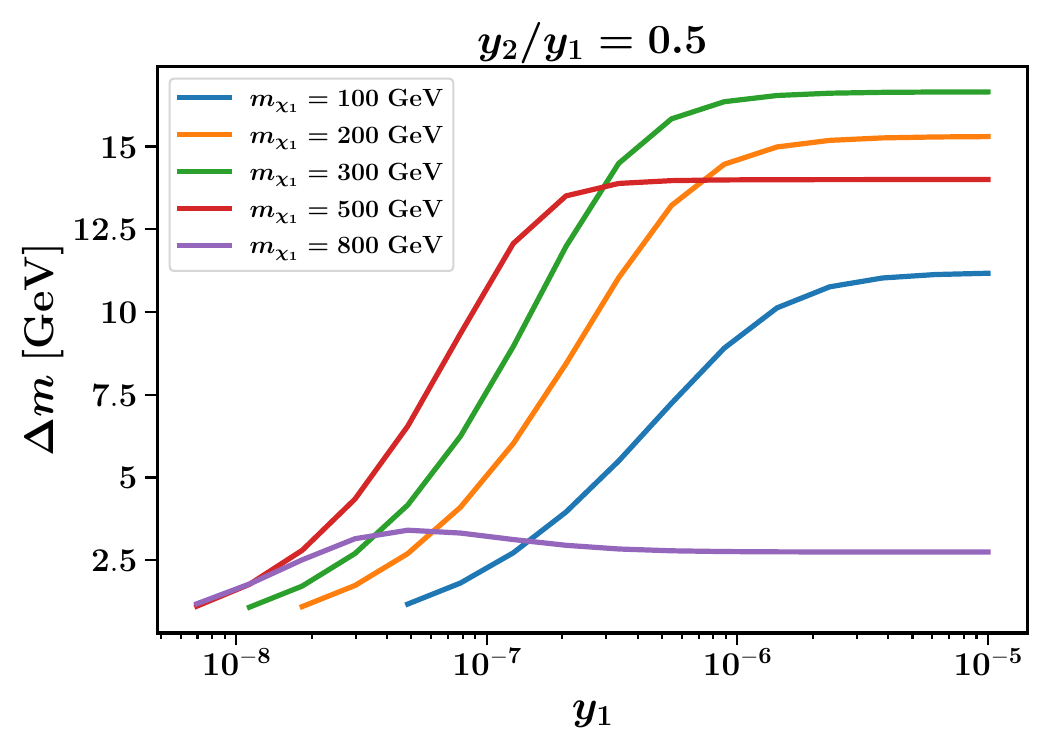}
    \end{subfigure}
    \hfill
    \begin{subfigure}[t]{0.48\textwidth}
        \centering
        \includegraphics[width=\linewidth]{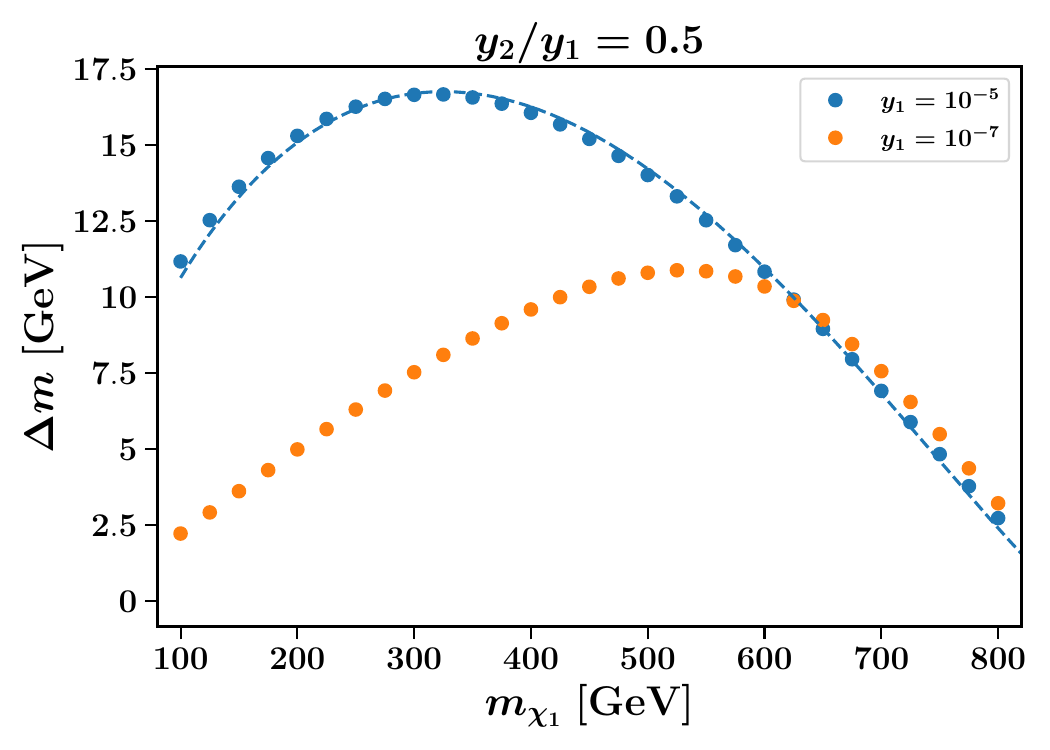}
    \end{subfigure}

    \caption{Mass splitting $\Delta m \equiv M_D-M_S$ along the $\Omega_\chi h^2=0.12$ contour. Left: $\Delta m$ as a function of $y_1$ for different values of the dark matter mass $m_{\chi_{1}}$. Right: the mass splitting as a function of $m_{\chi_{1}}$ in the co-annihilation regime, $y_1=10^{-5}$, and in the co-scattering regime, $y_1=10^{-7}$. The dots denote the data while the dashed curve is the fit from Eq.~\eqref{eq:CAfitfunction},  valid in the co-annihilation regime.}
    \label{fig:deltaM_comparison}
\end{figure}

Next, we discuss the mass splitting $\Delta m$ necessary to realize the thermal relic abundance.  This quantity controls the relative Boltzmann-suppression of the singlet and doublet sectors, and so is crucial to set effective co-annihilation and co-scattering rates.  This quantity is also of  interest because it impacts potential collider searches, see Sec.~\ref{sec:Signatures}.  The mass splitting  $\Delta m$ that  reproduces the observed relic abundance as a function of $y_1$ is shown in Fig.~\ref{fig:deltaM_comparison} (left panel) for  a fixed ratio of Yukawa couplings $y_2/y_1=0.5$. The curves exhibit the following general behavior: for sufficiently large $y_1$, $\Delta m$ approaches a constant value, while at smaller $y_1$, the required mass splitting decreases as the Yukawa coupling is reduced. This can be understood as follows: for $y_1 \gtrsim 10^{-5}$, all values of $M_S$ shown lie well within the co-annihilation region of Fig.~\ref{fig:parameterSpace}. The Yukawa couplings are large enough to maintain chemical equilibrium between the dark matter and the doublet sector, but otherwise serve little purpose: the relic abundance is primarily controlled by the part of $\langle\sigma v_{eff}\rangle$ that is proportional to  $\langle\sigma v_{\sect{2}\sect{2}\sect{0}\sect{0}}\rangle$ (see Eq.(\ref{eq:sigmaVeff})).  Both $\langle\sigma v_{\sect{2}\sect{2}\sect{0}\sect{0}}\rangle$ and the effective Boltzmann suppression
 depend on $M_D$ but are nearly independent of the Yukawa couplings. This is the origin of the $\Delta m$ plateau at large values of $y_{1}$.  
As the Yukawa couplings decrease and the system enters the co-scattering region of Fig.~\ref{fig:parameterSpace}, the value of $\Delta m$ required to reproduce the dark matter abundance decreases as the Yukawa coupling decreases.  This can be understood by considering the dominant contribution to the co-scattering amplitude which, as discussed above,  (for  $y_2/y_1\neq -1$)  depends on a $\chi^0_{2}-\chi^0_{1}-V$ coupling scaling as $\sim v y_+/\Delta m$. Therefore, as the Yukawa coupling decreases, a smaller mass splitting is required to maintain a sufficiently large co-scattering rate and reproduce the observed relic abundance. 

The value of the $\Delta m$ plateau reached at $y_1 \simeq 10^{-5}$  (see right edge of left panel) exhibits a nontrivial dependence on the dark matter mass. 
This dependence is explicitly illustrated in the right panel, see also, e.g. Ref.~\cite{Bhattiprolu:2025beq}.  The plateau is an initially increasing function of $M_S$, reaches a maximum near $M_S\simeq 300~\mathrm{GeV}$, and then decreases. We now present a simple argument to understand this shape semi-analytically. When the singlet and doublet states remain in chemical equilibrium throughout freeze-out,  the evolution of the total abundance is well-described by the single differential equation Eq.~(\ref{eq:YODE}) with effective cross section $\langle \sigma v_{eff} \rangle$ given as Eq.~(\ref{eq:sigmaVeff}). The contribution to the effective annihilation rate is dominated  by $\langle \sigma v_{\sect{2}\sect{2}\sect{0}\sect{0}}\rangle$, the thermally averaged annihilation cross section associated with a pure doublet multiplet, which is predominantly $s$-wave. 

The  dark matter relic abundance may be approximately determined in terms of this effective cross section. An approximation to the abundance can be found by integrating the effective cross section in the usual way \cite{Kolb:1990vq} from freeze-out $x_{f}$ to $\infty$. The result is that the relic abundance is approximately inversely-proportional to 
\begin{equation}\label{eq:CAfitfunction}
    J
    =
    \frac{A}{(M_S+\Delta m) ^2}
    \int_{x_f}^{\infty}
    \frac{dx}{x^2}
    \left(\frac{q(x)}{1+q(x)}\right)^2,
\end{equation}
where
\begin{equation}
q(x)
\equiv
\frac{n_\sect{2}^{\rm eq}}
{n_\sect{1}^{\rm eq}}
\approx
4
\left(1+\frac{\Delta m}{M_S}\right)^{3/2}
\exp\left(-x\frac{\Delta m}{M_S}\right).
\end{equation}
Because there is a fixed value of $J$ that leads to the correct relic density, this equation implicitly determines the mass splitting $\Delta m$ as a function of the dark matter mass $M_S$. This implicit equation gives the dashed curve for $\Delta m$ as a function of $m_{\chi_{1}}$ shown in Fig.~\ref{fig:deltaM_comparison} (right).\footnote{The overall constant of proportionality that can be extracted from the pure doublet ($\Delta m \rightarrow 0$) limit.}  The dots are points that yield the correct relic density by direct numerical integration of the Boltzmann equation.  The agreement is excellent.  On the other hand, for $y_{1}= 10^{-7}$, the correct relic abundance is instead realized for the orange dots in the right panel.  This value of the Yukawa coupling is deep in the co-scattering regime; the departure of these from the dashed blue curve shows the inadequacy of the simple 1-ODE co-annihilation picture.

\begin{figure}[ht!]
    \centering
    \begin{subfigure}{0.48\textwidth}
        \centering
        \includegraphics[width=\linewidth]{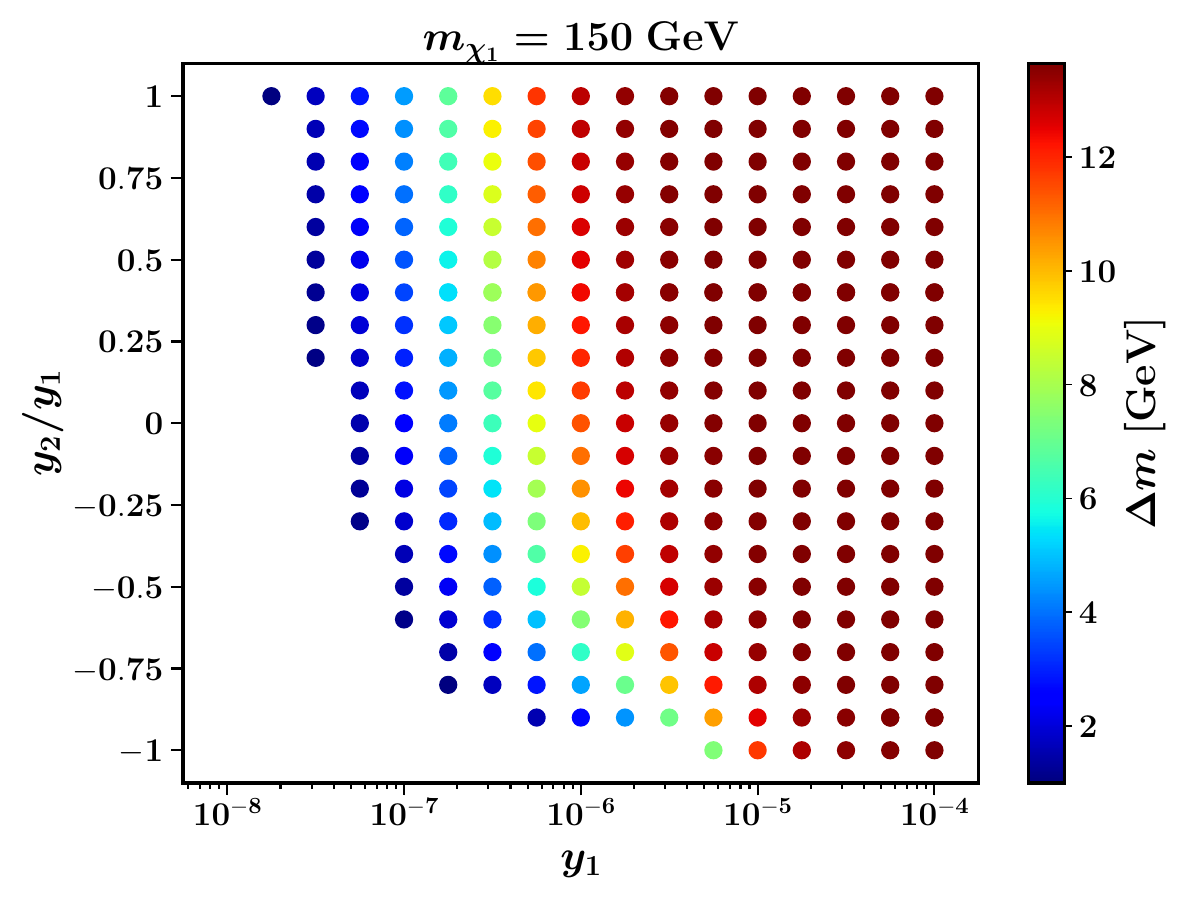}
    \end{subfigure}
    \hfill
    \begin{subfigure}{0.48\textwidth}
        \centering
        \includegraphics[width=\linewidth]{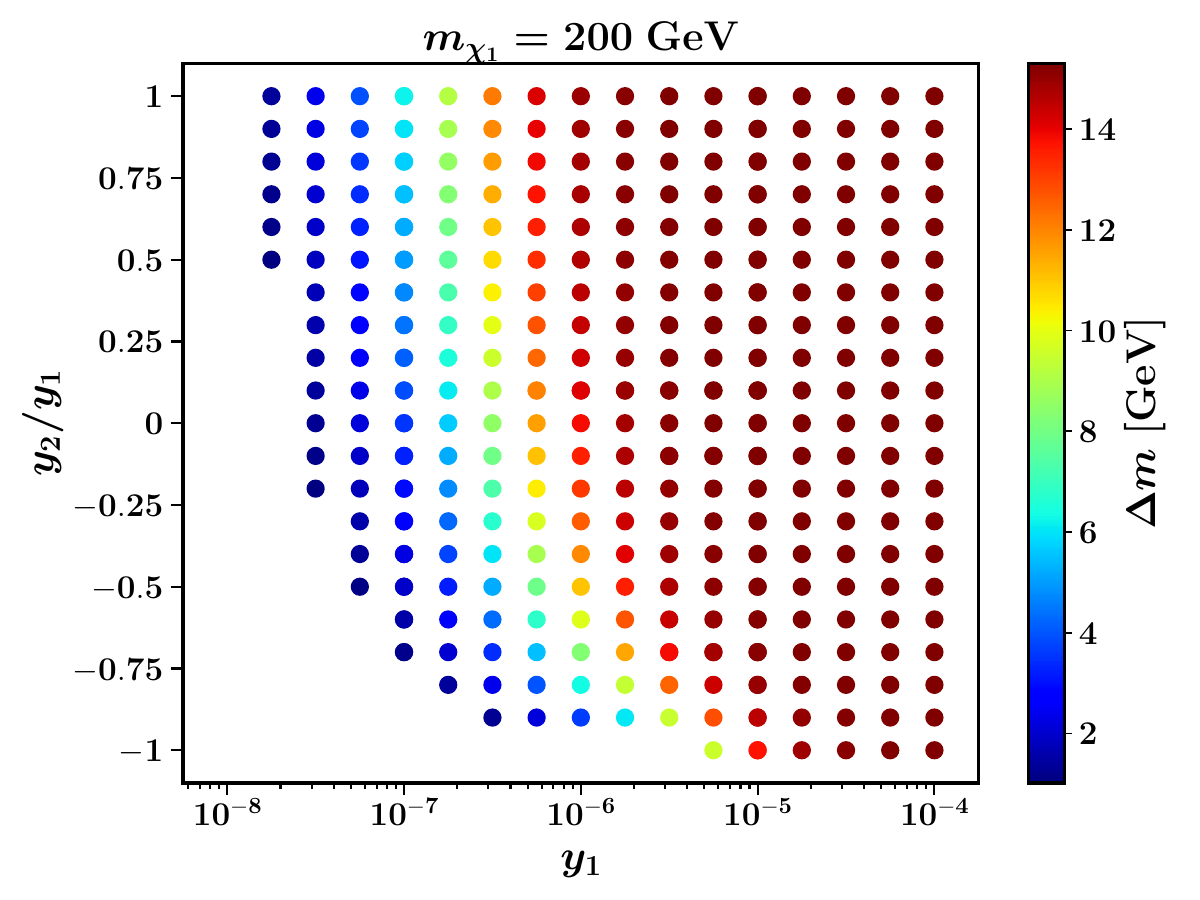}
    \end{subfigure}

    \begin{subfigure}{0.48\textwidth}
        \centering
        \includegraphics[width=\linewidth]{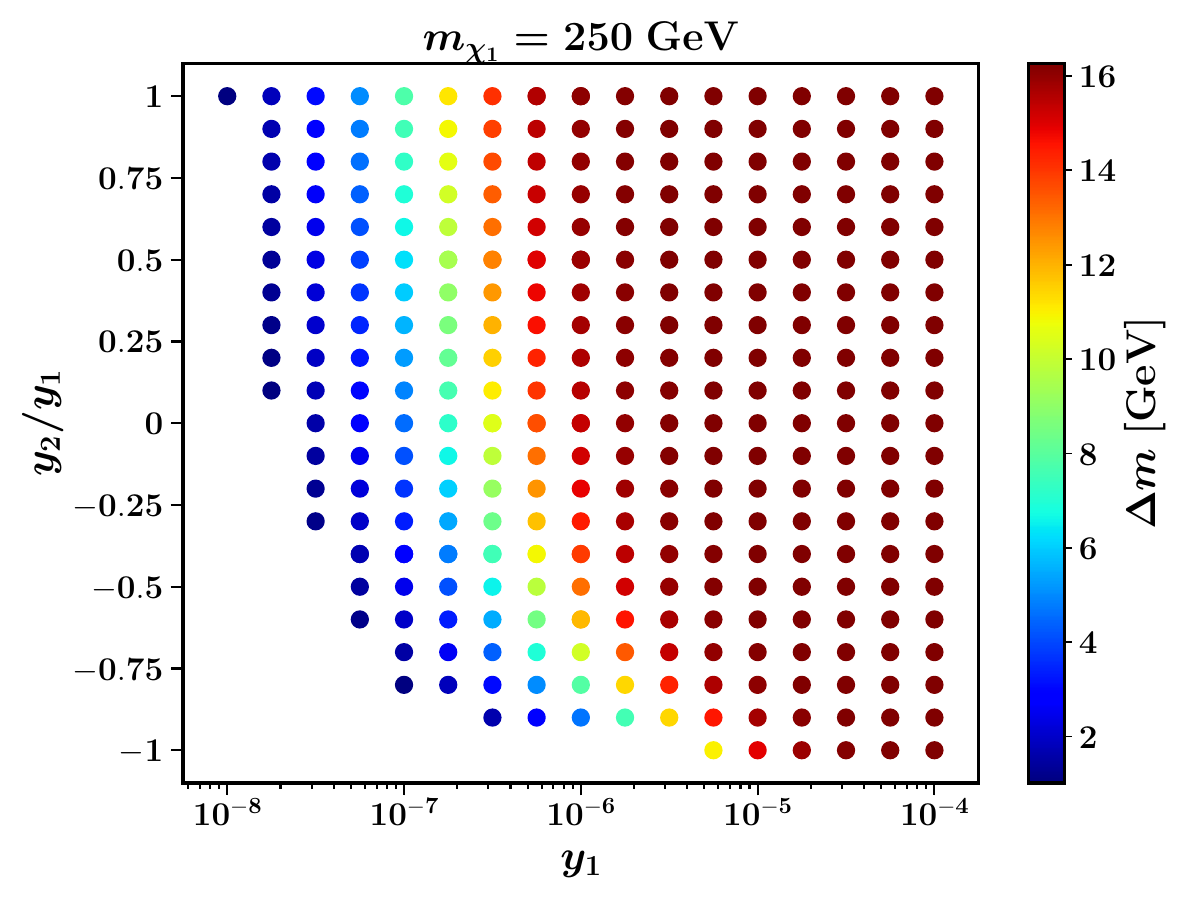}
    \end{subfigure}
    \hfill
    \begin{subfigure}{0.48\textwidth}
        \centering
        \includegraphics[width=\linewidth]{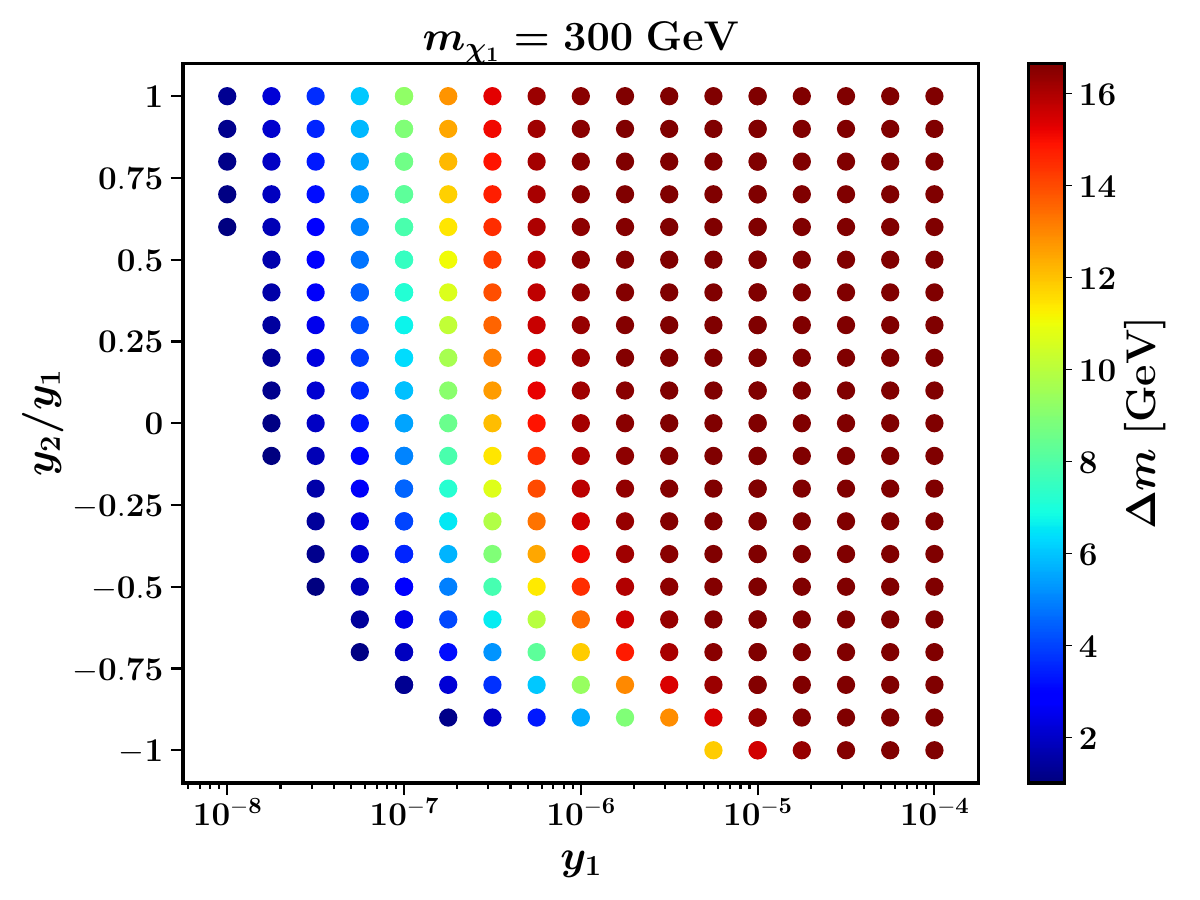}
    \end{subfigure}

    \caption{The color bar indicates the mass splitting $\Delta m=M_D-M_S$ required to obtain $\Omega h^2=0.12$ through freeze-out production in the $(y_1,y_2/y_1)$ parameter space. The four panels correspond to fixed dark matter masses $m_{\chi_1}=150$, $200$, $300$, and $250~\mathrm{GeV}$ (clockwise from upper left).}
    \label{fig:ratio_vs_y1}
\end{figure}

Having understood the size of the mass splitting for fixed $y_{2}/y_{1}$, we now turn to the dependence of $\Delta m$ on this ratio. This is shown in Fig.~\ref{fig:ratio_vs_y1} for four representative values of $m_{\chi_1}$, where the mass splitting required to reproduce the observed relic abundance is displayed in the ($y_1$,$y_2/y_1$) plane. Again, for $y_1$ sufficiently large to maintain chemical equilibrium between the dark matter and doublet states, the required value of $\Delta m$ is largely insensitive to both $y_1$ and $y_2/y_1$. This can be seen near $y_1\sim10^{-4}$.  For fixed $y_2/y_1$, decreasing $y_1$ requires a corresponding decrease in $\Delta m$ to maintain $\Omega h^2=0.12$, as discussed above. 
Now, at fixed $y_1$, the required mass splitting generally decreases as $y_2/y_1$ is varied from $1 \rightarrow -1$. These statements follow from the general scaling for co-scattering $y_+v/\Delta m$ when $y_2/y_1\neq-1$. Points with $\Delta m<1~\mathrm{GeV}$ are not displayed.

\subsection{Freeze-In regime}
\label{sec:FIRegime}

\begin{figure}
    \centering

    \begin{subfigure}[t]{0.49\linewidth}
        \centering
        \includegraphics[width=\linewidth]{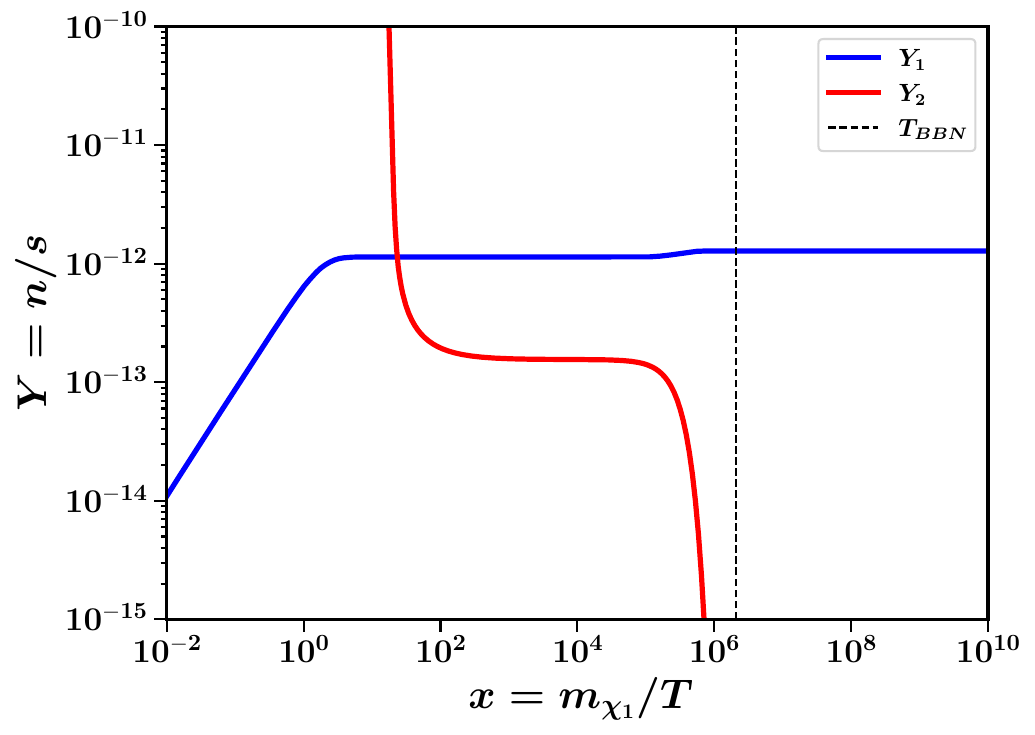}
    \end{subfigure}
    \hfill
    \begin{subfigure}[t]{0.49\linewidth}
        \centering
        \includegraphics[width=\linewidth]{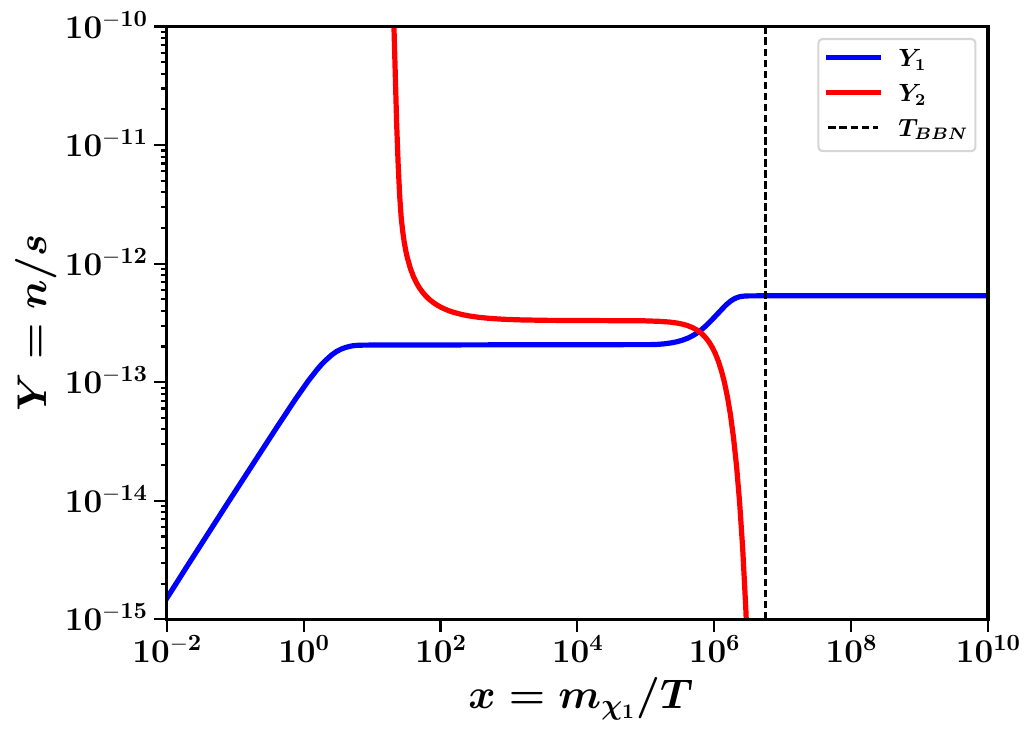}
    \end{subfigure}

    \caption{Evolution of the comoving dark matter and doublet abundances, $Y_\sect{1}$ and $Y_\sect{2}$, during freeze-in production. Left panel: ($M_S=300$ GeV, $M_D=400$ GeV, $y_1=9.82\times10^{-12}$, $y_2=4.91\times10^{-12})$;   right panel:  
($M_S=800$ GeV, $M_D=900$ GeV, $y_1=5.90\times10^{-12}$, $y_2=2.95\times10^{-12}$). At late times, the doublet states decay into dark matter, increasing $Y_\sect{1}$. The vertical dashed lines mark the temperature associated with Big Bang nucleosynthesis, corresponding to a cosmic time of $\tau \sim 50~\mathrm{s}$.}
    \label{fig:YvsXFI}
\end{figure}

We now discuss the build up of dark matter via freeze-in. Here, the dark matter is never in chemical equilibrium with the doublet sector and thus never in thermal equilibrium with the SM bath. We 
use Eq.~(\ref{eq:freezeinevolution}) accounting for one-directional processes of decay of doublet states, SM-annihilations, and co-scattering. We include the effects of the thermal masses of electroweak bosons via a cut on the transverse momentum $p_T$ on the $t$-channel integral when calculating the cross section, following  \cite{Alguero:2023zol}.   The contribution of early universe decays (when the dark sector is at its full $\propto T^3$ abundance) to the dark matter relic abundance can be solved analytically for the singlet-doublet model \cite{Calibbi:2018fqf} yielding:
\begin{equation}
\label{eq:FIAnalytic}
\Omega_{FI}^{decay} h^2
=
0.1
\left(\frac{105}{g_*}\right)^{3/2}
\left(\frac{M_S}{100~\mathrm{GeV}}\right)
\left(\frac{1~\mathrm{TeV}}{M_D}\right)^2
\left(
\frac{\sum_{\psi_i\in\sect{2}} g_{\psi_i}\Gamma_{\psi_i \rightarrow \sect{1}, \sect{0}}}
{5\times 10^{-22}~\mathrm{GeV}}
\right),
\end{equation}
where $g_*$ is the effective number of relativistic degrees of freedom evaluated at $T\sim M_D$. Here, the sum runs over the doublet sector particles ($\chi^0_{2}, \chi^0_{3}, \chi^{\pm}$) so
$\sum_{\psi_i\in\sect{2}} g_{\psi_i}\Gamma_{\psi_i \rightarrow \sect{1}, \sect{0}}$
represents the total decay rate from the doublet sector to the dark matter $\chi^0_1$ and the SM.   
This expression does not account for very late decays of doublet particles after their freeze-out, known as the SuperWIMP mechanism \cite{Feng:2003uy, Feng:2003xh}. This contribution is given by the approximate expression:
\begin{equation} \label{eq:superwimp}
    \Omega_{SW} h^2 \approx 0.1\left (\frac{M_{D}}{\text{1 TeV}} \right)^2 \frac{M_{S}}{M_{D}}.
\end{equation}
The coefficient and first ratio are the would-be relic density of the doublet sector.  The last factor results from the difference in the doublet and singlet masses.  The SuperWIMP contribution becomes more significant for larger values of the doublet mass $M_D$.  As we will see below, the relative contribution to freeze-in from co-scattering processes is especially significant at small mass splittings $\Delta m \lesssim 100$ GeV, in which case Eq.~(\ref{eq:FIAnalytic}) and Eq.~(\ref{eq:superwimp}) are insufficient to describe the freeze-in, and we must rely on the numerical evaluation.

The evolution of the comoving abundances $Y_\sect{1}$ and $Y_\sect{2}$ is shown in Fig.~\ref{fig:YvsXFI} for two choices of parameters:   $M_S=300$ GeV, $M_D=400$ GeV, $y_1= 9.82\times 10^{-12}$,
$y_2=4.91 \times 10^{-12}$ (left) and   $M_S=800$ GeV, $M_D=900$ GeV, $y_1= 5.90 \times 10^{-12}$,
$y_2=2.95 \times 10^{-12}$ (right).  For both sets of parameters, at early times, $x\lesssim 1$, the doublet states remain in thermal equilibrium and are relativistic, while $Y_\sect{1}$ is gradually populated through freeze-in. As the temperature falls below the doublet mass, the equilibrium abundance of the doublet states decreases exponentially. Consequently, freeze-in production becomes inefficient and $Y_\sect{1}$ approaches an initial plateau, while $Y_\sect{2}$ eventually departs from equilibrium and freezes out. Following the freeze-out of the doublet states, the co-moving abundances remain approximately constant until the remaining doublet states decay at much later times. This late-time transfer is the SuperWIMP contribution. In the left panel of Fig.~\ref{fig:YvsXFI} this is a small correction to the earlier freeze-in abundance, whereas it constitutes the dominant contribution to the final $\chi^0_1$ abundance in the right panel. In both cases, the doublet states decay before $T_{\mathrm{BBN}}$, and the resulting final abundance is $\Omega h^2=0.12$.
\begin{figure}[t]
    \centering

    \begin{subfigure}{0.48\textwidth}
        \centering
        \includegraphics[width=\linewidth]{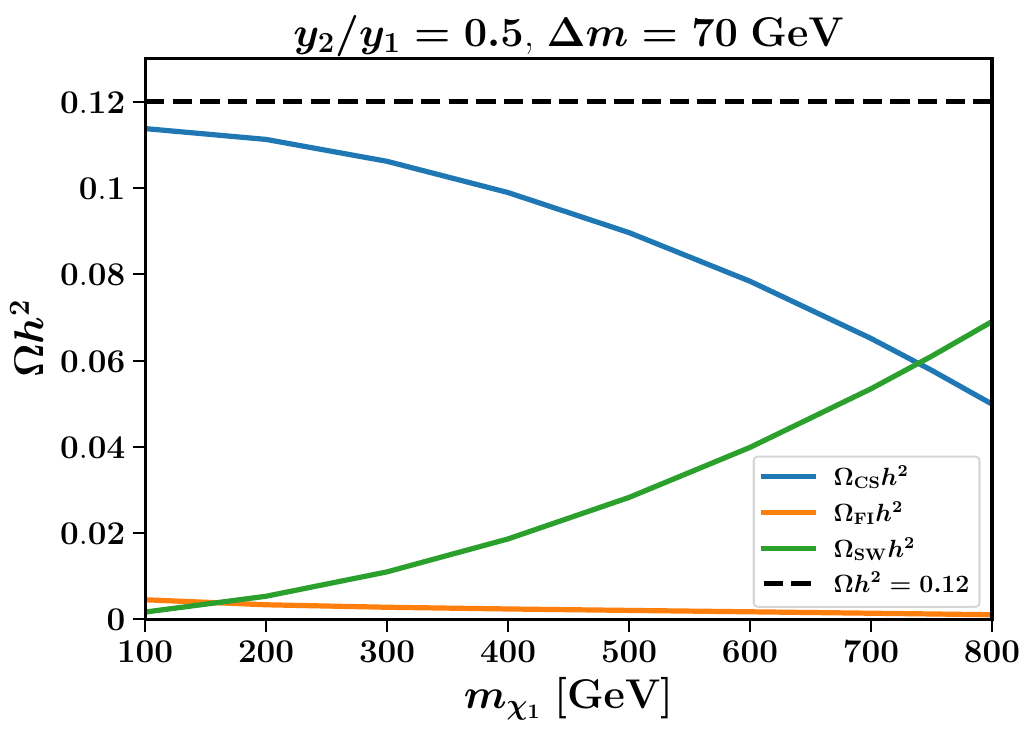}
    \end{subfigure}
    \hfill
    \begin{subfigure}{0.48\textwidth}
        \centering
        \includegraphics[width=\linewidth]{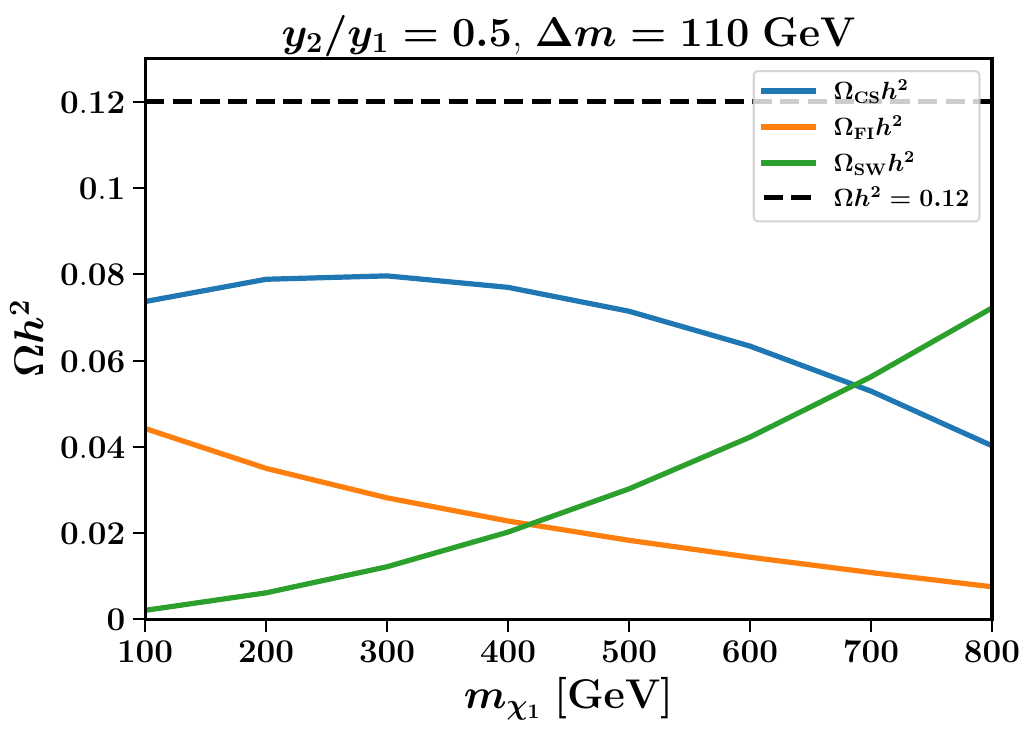}
    \end{subfigure}

    \begin{subfigure}{0.48\textwidth}
        \centering
        \includegraphics[width=\linewidth]{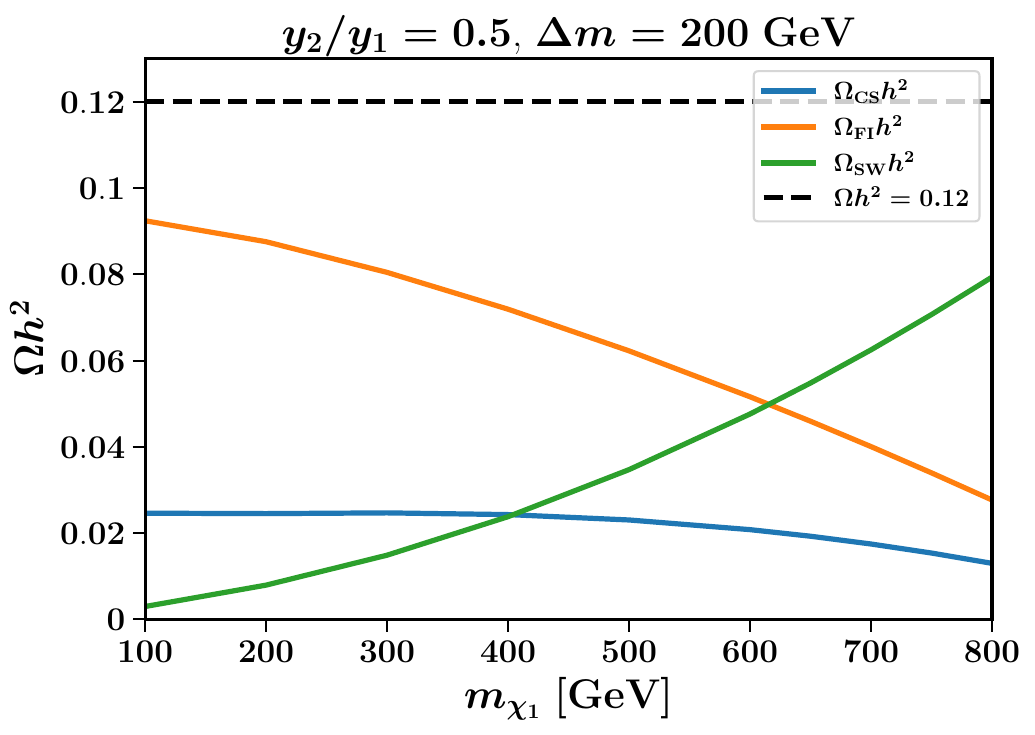}
    \end{subfigure}
    \hfill
    \begin{subfigure}{0.48\textwidth}
        \centering
        \includegraphics[width=\linewidth]{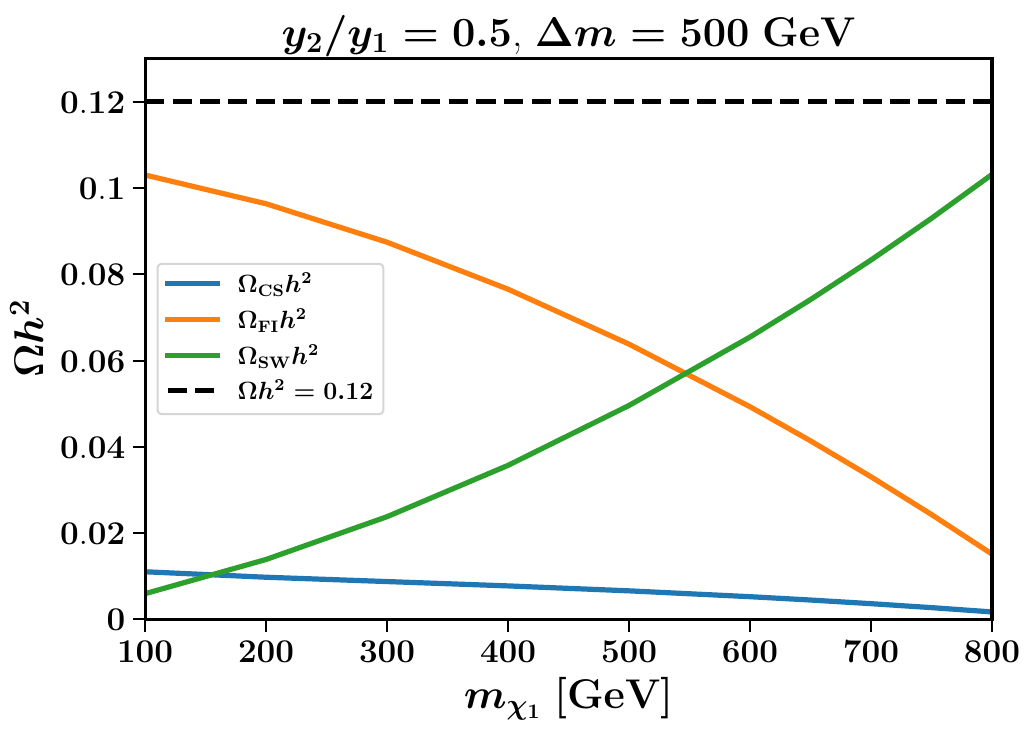}
    \end{subfigure}

    \caption{Contributions to the dark matter relic density $\Omega h^2=0.12$ via freeze-in from decays and SM annihilations $\Omega_{FI}h^2$, co-scattering $\Omega_{CS}h^2$, and late decays $\Omega_{SW}h^2$ as a function of dark matter mass $m_{\chi_1}$.  The mass splitting between the doublet and singlet mass is $\Delta m = 70, 110, 500, 200$ GeV (clockwise from upper-left).  For all panels,  the ratio of Yukawa couplings $y_{2}/y_{1}= 0.5$, and  $y_{1}$ is fixed so that the total abundance matches the observed value.}
    \label{fig:Omegah2contribution}
\end{figure}

The relative importance of decays, SM annihilations, and co-scattering is controlled primarily by the mass splitting, whereas the SuperWIMP contribution depends mainly on the dark matter mass. This behavior is illustrated in Fig.~\ref{fig:Omegah2contribution}, where the relic abundance is decomposed as $\Omega h^2=\Omega_{\mathrm{FI}}h^2+\Omega_{\mathrm{CS}}h^2+\Omega_{\mathrm{SW}}h^2=0.12.$ In this Figure, $\Omega_{\mathrm{FI}}h^2$ includes freeze-in production from doublet decays and SM annihilations, $\Omega_{\mathrm{CS}}h^2$ denotes the co-scattering contribution, and $\Omega_{\mathrm{SW}}h^2$ arises from the late decays of the frozen-out doublet population (the SuperWIMP contribution). At each value of $M_S$, the coupling $y_{1}$ is chosen so that the total relic abundance satisfies $\Omega h^2=0.12$. While SM annihilation processes are included in the calculation, they are subdominant to decays for all points considered, typically contributing less than $10\%$ of the freeze-in abundance. 
The shape of the $\Omega_{\mathrm{SW}}h^2$ curve follows directly from Eq.~(\ref{eq:superwimp}).
Setting $\Omega_{SW} h^2=0.12$, and writing $M_{D} = M_{S} + \Delta m$, there is a critical value of singlet mass $M_{S} = m_{c}$, which causes the SuperWIMP contribution to comprise the entire observed relic abundance.
\begin{equation}
\label{eq:cutoffmass}
    m_{c}=\frac{-\Delta m+\sqrt{\Delta m^2+\left(2200 \, {\rm GeV}\right)^2}}{2}
    ,
\end{equation}

Below, we discuss the relative contributions from co-scattering, freeze-in decays, and the SuperWIMP contribution for different values of $\Delta m$ and dark matter mass as illustrated in Fig.~\ref{fig:Omegah2contribution}.
For $\Delta m=70~\mathrm{GeV}$ (upper left), co-scattering provides the dominant contribution over most of the dark matter mass range, while decays are negligible. This is because two-body decays are kinematically forbidden and the remaining three-body decay widths are suppressed. At the largest dark matter masses shown the SuperWIMP contribution becomes comparable to and eventually exceeds the co-scattering contribution.
For $\Delta m=110~\mathrm{GeV}$ (upper right), co-scattering remains the largest contribution over most of the parameter range, although decays make a non-negligible contribution at small $M_S$. While some two-body decay channels are  kinematically accessible for this mass splitting, the limited available phase space still suppresses the decay widths. The SuperWIMP contribution again becomes increasingly important at large $M_S$. At $\Delta m=200~\mathrm{GeV}$ and $\Delta m=500~\mathrm{GeV}$ (lower panels) the two-body decays are no-longer phase-space suppressed, so the freeze-in contribution is substantially enhanced, and dominates the co-scattering contribution. 

\begin{figure}[t]
    \centering

    \begin{subfigure}{0.48\textwidth}
        \centering
        \includegraphics[width=\linewidth]{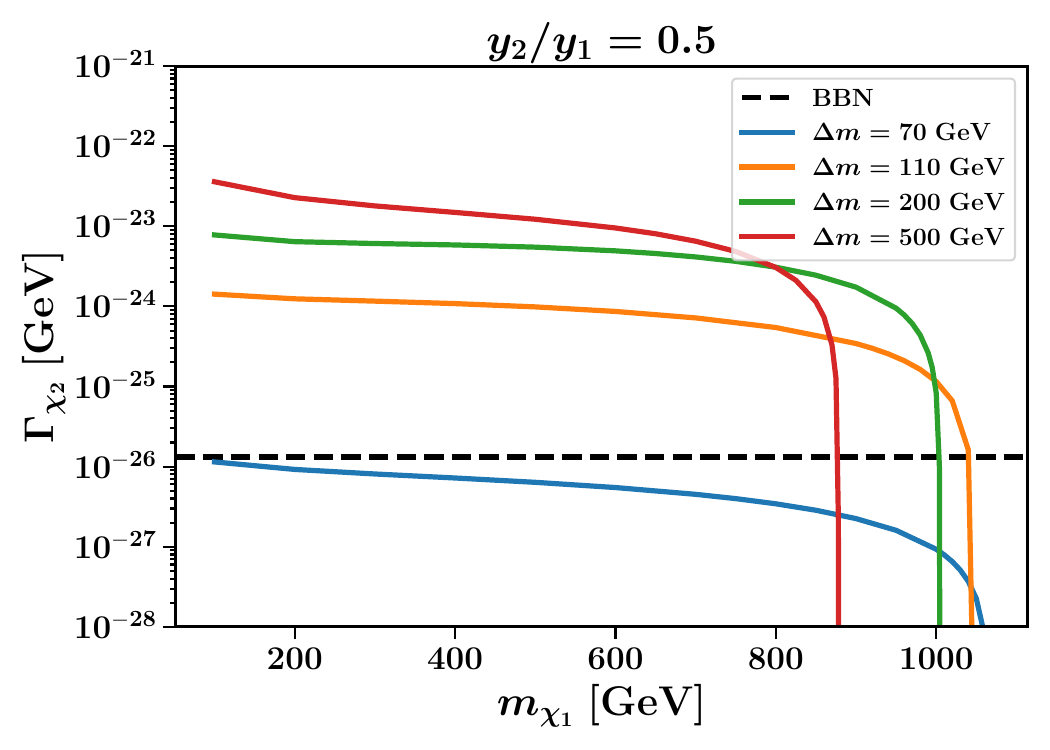}
    \end{subfigure}
    \hfill
    \begin{subfigure}{0.48\textwidth}
        \centering
        \includegraphics[width=\linewidth]{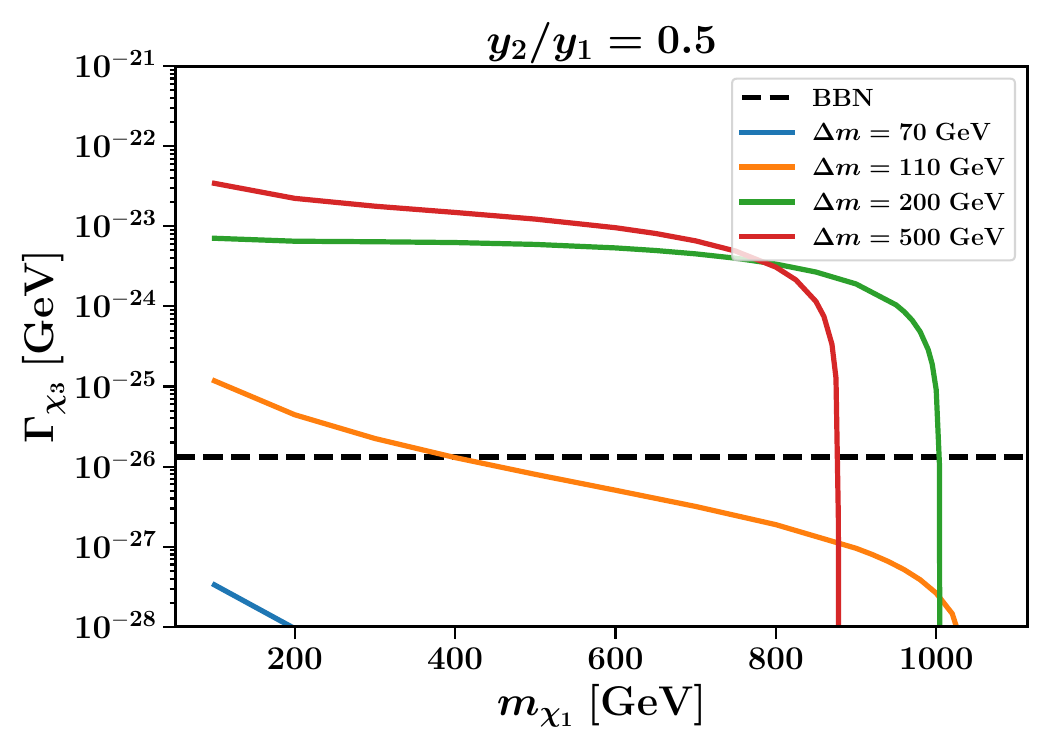}
    \end{subfigure}

    \vspace{0.35cm}

    \begin{subfigure}{0.48\textwidth}
        \centering
        \includegraphics[width=\linewidth]{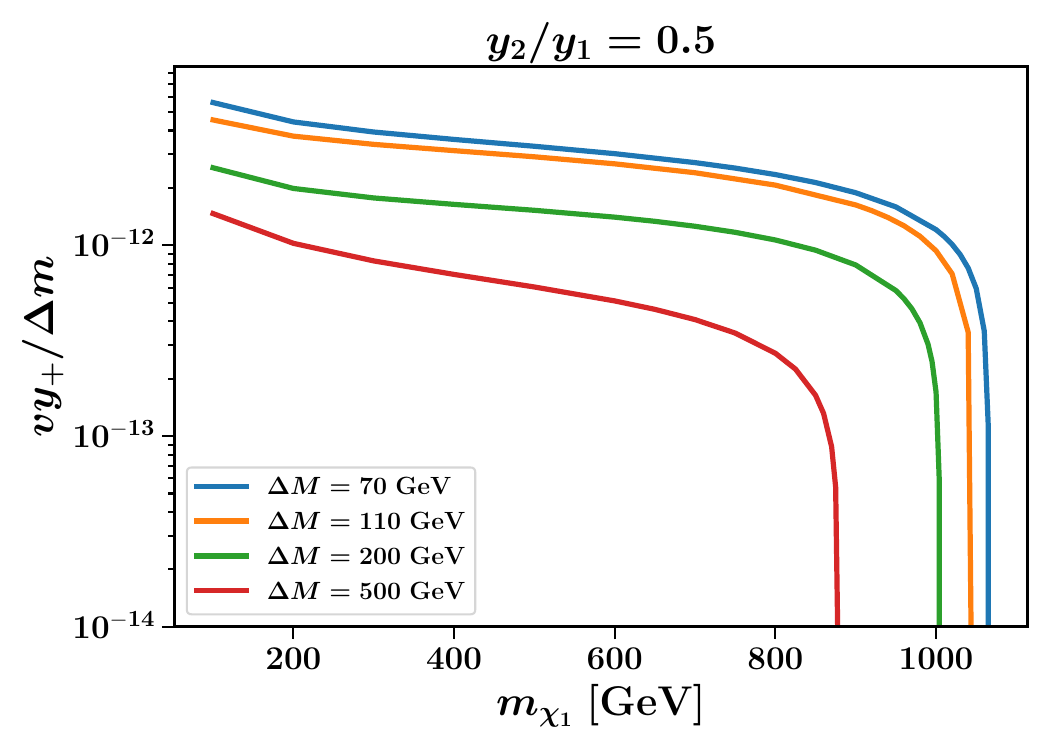}
    \end{subfigure}

    \caption{Top panels: Total Decay widths $\Gamma_{\chi_i}$ for $i=2$ (left) and $i=3$ (right) as a function of the dark matter mass $m_{\chi_1}$ for parameter points satisfying $\Omega_\chi h^2=0.12$ through freeze-in production. The dashed black line indicates the bound from Big Bang nucleosynthesis, corresponding to a lifetime of $\tau \sim 50~\mathrm{s}$. Bottom panel: Corresponding values of $y_{+}$.}
    \label{fig:Decay0.5}
\end{figure}

For fixed values of $y_{2}/y_{1}$, $\Delta m$ and $m_{\chi_1}$, enforcing that the dark matter reproduce the observed relic abundance uniquely determines the  total decay widths of $\chi^0_2$ and $\chi^0_3$, denoted $\Gamma_{\chi_2}$ and $\Gamma_{\chi_3}$ respectively. 
These decay widths are shown in Fig.~\ref{fig:Decay0.5}. If the doublet lifetimes are too long, they have the potential to disturb the successful predictions of BBN.  We indicate the approximate BBN constraint \cite{Kawasaki_2018} by requiring a lifetime $\tau\lesssim 50~\mathrm{s}$, corresponding to $\Gamma\gtrsim 1.3\times10^{-26}~\mathrm{GeV}$. The precise BBN bound depends on both the abundance of the decaying state and the visible decay products, so this criterion should be regarded as an approximate but accurate benchmark.

At the temperatures corresponding to BBN the doublet states $\chi^0_2$, $\chi^0_3$, and $\chi^{\pm}$ are no longer in chemical equilibrium. By this time, the charged states $\chi^{\pm}$ have decayed to $\chi^0_2$/$\chi^0_3$ accompanied by $\pi^{\pm}$, with equal branching fractions, as shown in Eq.~\eqref{eq:Cha_to_chi2_pion}. Thus, around the time of BBN, the remaining doublet sector yield is composed of approximately half $\chi^0_2$ and half $\chi^0_3$. Both $\chi^0_2$ and $\chi^0_3$ must decay before the BBN epoch. 
In the three-body regime, the decay width strongly depends on the available phase space. For mass splittings near the upper end of the three-body regime ($\Delta m = 70-90$ GeV), $\chi^0_2$ could potentially decay before BBN (left panel). However, $\chi^0_3$ remains long-lived enough to violate the BBN bound (right panel). 
The coupling of the $\chi^0_{3}$ to the $Z$ boson and $\chi^0_{1}$ is substantially smaller than the corresponding coupling for $\chi^0_{2}$.  Although there may sometimes be a $\chi^0_{3}-\chi^0_{1}-h$ coupling larger than the coupling to the $Z$ boson, as long as $\Delta m < m_{h}$, the $\chi^0_{3}$ lifetime is substantially suppressed. This behavior is illustrated for $\Delta m = 110$ GeV, where $\chi^0_2$ decays well before BBN, while $\chi^0_3$ only narrowly satisfies the BBN constraint for $m_{\chi_1}\lesssim 400$ GeV. 
Once two-body decays into on-shell Higgs become kinematically accessible, the $\chi^0_3$ decays well before the BBN era. Importantly, although the $\chi^0_{3}$ partial width to $\chi^0_{1}+Z$ is suppressed by $\Delta m/(M_S+M_D)$, the partial width to $\chi^0_{1}+h$ dominates \cite{Calibbi:2018fqf}, leading to comparable decay widths for $\chi^0_2$ and $\chi^0_3$. As a result, $\Gamma_{\chi_2}\approx\Gamma_{\chi_3}$ for mass splittings of 200 and 500 GeV.

 The rapid drop in decay widths observed at large values of $m_{\chi_{1}}$ is due to the presence of the SuperWIMP contribution to the relic density.
From Eq.~(\ref{eq:superwimp}), the SuperWIMP contribution increases with $M_S$ at fixed $\Delta m$, independent of the Yukawa couplings. As $M_S$ increases, a smaller contribution is required from the (coupling-dependent) freeze-in and co-scattering processes.  Eventually, the SuperWIMP contribution saturates the relic abundance (see Eq.~(\ref{eq:cutoffmass})). In this case, there is no room for a freeze-in contribution.  The bottom panel of Fig~\ref{fig:Decay0.5} shows the corresponding Yukawa values. The approximate value of the Yukawa couplings to obtain the observed relic abundance via freeze-in is $y_1\sim y_2\sim10^{-12}$.

\section{Experimental Signatures}
\label{sec:Signatures}

In the small-Yukawa-coupling regime, direct dark matter couplings to the SM are negligible.  Because of this, both indirect and direct detection signals are strongly suppressed.  For example, direct detection signals will be dominated by the loop-induced contribution \cite{Chen:2019gtm, Chen:2018uqz}, and will lie well below the neutrino fog. 

The best hope for discovery will be detection of the doublet sector.  In the following, we review relevant collider studies and make comments on additional searches that might be done at the LHC and beyond.  We plan a more detailed collider analysis in future work \cite{Takla:ToAppear}.

\begin{figure}
    \centering

    \begin{subfigure}[t]{0.49\linewidth}
        \centering
        \includegraphics[width=\linewidth]{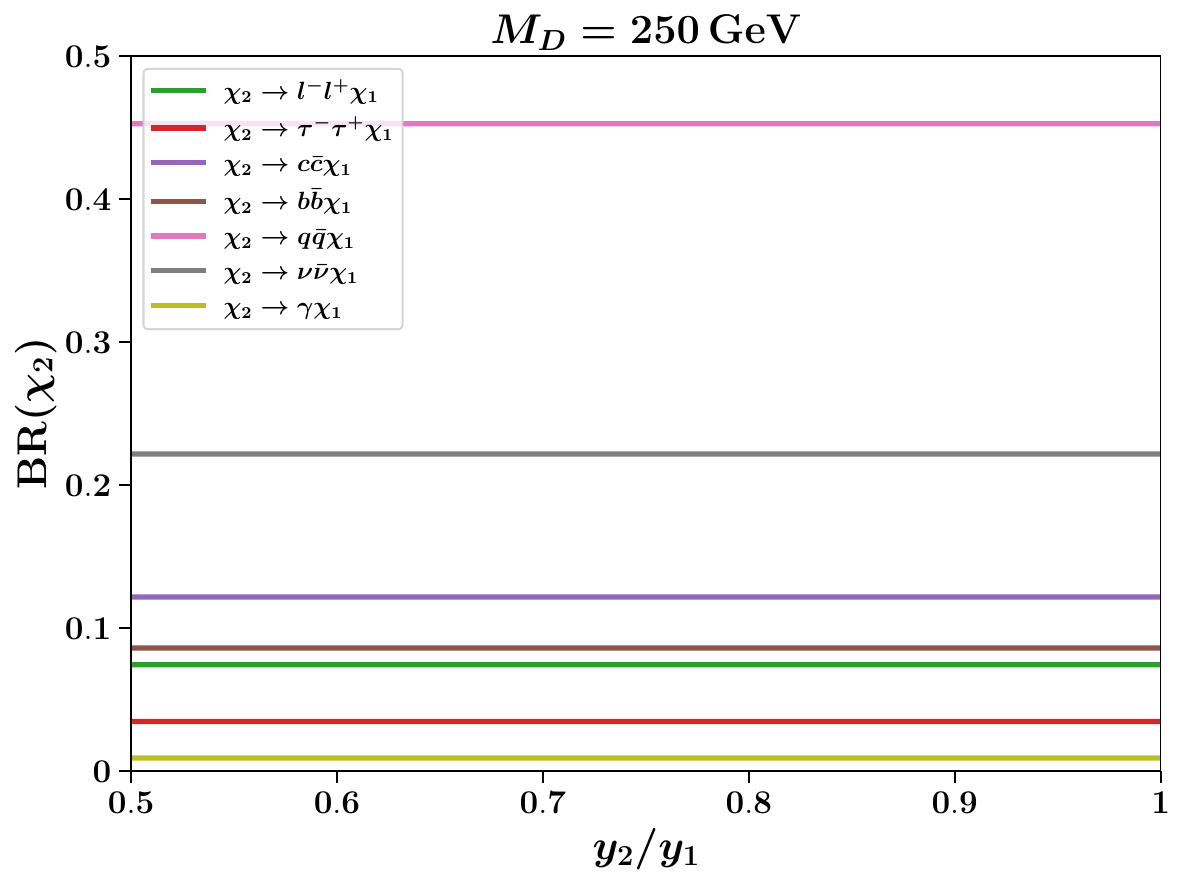}
    \end{subfigure}
    \hfill
    \begin{subfigure}[t]{0.49\linewidth}
        \centering
        \includegraphics[width=\linewidth]{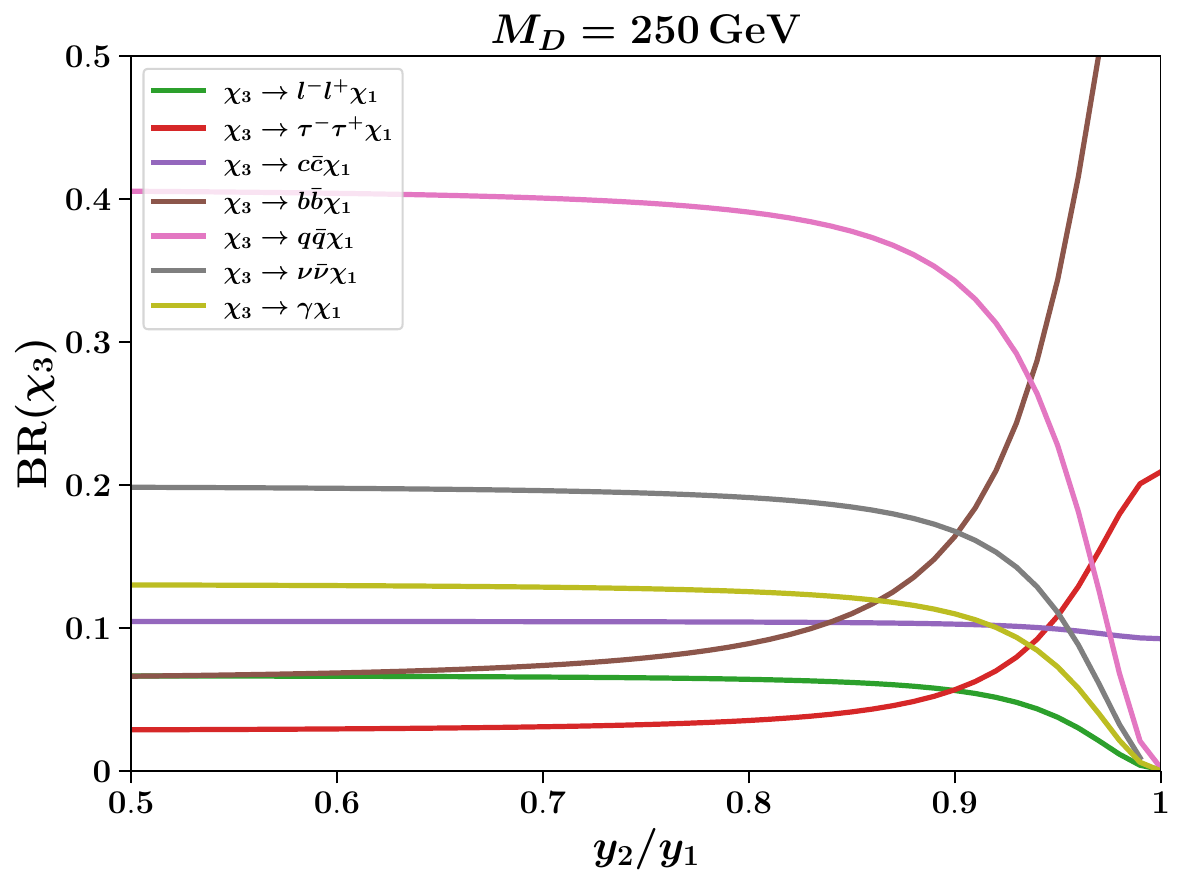}
    \end{subfigure}

    \caption{Branching ratio of $\chi^0_{2}$ (left) and $\chi^0_{3}$ (right) as a function of the ratio of Yukawa couplings $y_{2}/y_{1}$ for $M_{D}= 250$ GeV, $y_1=10^{-4}$ and $\Delta m=16$ GeV.  This value of $\Delta m$ reproduces the relic density for this $M_{D}$. The curve labeled $l^-l^+$ is the sum of the electron and muon contributions (each contributing half), the $q\bar{q}$ includes light quarks ($u$, $d$, $s$), $\nu\bar{\nu}$ is the sum of the three generations of neutrinos, and $\gamma$ is the loop induced radiative decay. The branching ratio for $\chi^0_{2}$ is independent of the $y_2/y_1$ as Z-mediated decays remain dominant throughout the entire parameter space. In contrast, the branching ratios of $\chi^0_{3}$ depend on $y_2/y_1$ because Higgs-mediated processes become competitive as $y_2/y_1\rightarrow 1$}
    \label{fig:BR}

\end{figure}

First, we note that production at colliders will be limited to the doublet states.  It is possible to produce pairs of charged states, a charged and a neutral state ($\chi^0_{2}$ or $\chi^0_{3}$), or a $\chi^0_{2}$ in association with a $\chi^0_{3}$.  All these states will be produced with electroweak cross section, which is ${\mathcal O}$(pb) for $M_{D} \approx 100 $ GeV, and is a rapidly falling function of mass, see, e.g. \cite{Bhattiprolu:2025beq}. We next discuss the branching ratio and lifetimes of these doublet states.  These will determine what decays can be relevant at colliders.

We first address the cases where $\Delta m \lesssim M_{W}$, as is relevant for the freeze-out regime.  In this case, two-body decays of the heavy dark states to $Z\chi^0_1$, $W \chi^0_1$ and $h \chi^0_1$ are all inaccessible.  In this case, three-body decays of the heavy dark sector states typically dominate. However, when the spectrum is compressed in this way, in some regions of parameter space, two-body decays from $\chi^0_{3} \rightarrow \chi^0_{1} \gamma$ can also be relevant. These decays are analogous to the radiative neutralino decays \cite{Haber:1988px,Baum:2023inl}.  The three-body decays obey the usual  $(\Delta m)^5$ scaling, while the radiative decays, while loop suppressed, scale as a lower power of $\Delta m$, so can compete.  We note decays between the neutral states $\chi^0_{3} \rightarrow \chi^0_{2} +X$ decays are strongly suppressed owing to the high degree of \ degeneracy between $\chi^0_{2}$ and $\chi^0_{3}$ in the small Yukawa limit.  

Expressions for three-body decays can be found by taking appropriate limits of the expressions for electroweakinos in Ref.~\cite{Djouadi:2001fa}.  We utilize the built-in functionality of Micromegas \cite{Alguero:2023zol} for numerical computations of the three-body partial decay widths, and find consistent results.  We modify expressions form \cite{Haber:1988px}, see also \cite{Baum:2023inl}, for the computation of radiative decays.  In Fig.~\ref{fig:BR} we display the branching ratios of $\chi^0_{2}$ and $\chi^0_{3}$  as a function of the ratio of the Yukawa couplings $y_{2}/y_{1}$ for a $M_{D} = 250$ GeV.  We have taken $y_{1}= 10^{-4}$ and $\Delta m= 16$ GeV, which reproduces the correct relic abundance for this choice of $M_{D}$ \footnote{The branching ratios are nearly identical for $10^{-6}<y_1<10^{-2}$. However, the $\chi^0_3\rightarrow\chi^0_2+X$ becomes important at the few percent level for $y_1\lesssim10^{-2}$.}. For all values of this ratio, we find three-body decays dominate, though radiative decays can contribute at the ${\mathcal O}(10\%)$ level for $\chi^0_3$. However, for $\chi^0_2$ it contributes at the  ${\mathcal O}(1\%)$ level. For $y_2/y_1\in[-1,0.5]$ the branching ratios are essentially constant and are therefore not shown. In this region, the decays are dominated by the $Z$-mediated contribution, and all the dominant partial widths depend on the same combination of the Yukawa couplings. In fact, for $\chi^0_2$ this is the case for all ratios, $y_2/y_1\in[-1,1]$. However, for $\chi^0_3$ as the $y_{2}/y_{1}$ ratio approaches unity, the relative importance of the diagram with an intermediate Higgs boson grows.  Diagrams with an intermediate Higgs boson are suppressed because they involve a small SM Yukawa coupling, in contrast to the diagram with an intermediate $Z$-boson that couples to the SM with the ${\mathcal O}(1)$ weak gauge coupling.  However, as $y_{2}/y_{1} \rightarrow 1$, the $Z -\chi^0_3-\chi^0_1$ coupling $\propto y_-$ vanishes,  only the $h-\chi^0_{3}-\chi^0_{1}$ coupling which is $\propto y_{+}$ remains relevant.  The increased importance of the off-shell Higgs boson diagram is reflected in the dominance of the $b \bar{b}$ and $\tau^{+} \tau^{-}$ final states near $y_{2}/y_{1} =1$.  

Which collider signals are relevant depends on the lifetime of the particles in the doublet sector.  For smaller Yukawa couplings, the decrease in decay rates for $\chi^0_{3} \rightarrow \chi^0_{1} f \bar{f}$ and $\chi^0_{2} \rightarrow \chi^0_{1} f \bar{f}$ can lead to displaced vertices.  The lifetimes of the $\chi^0_{2}$ and $\chi^0_{3}$ states are controlled by different combinations of the Yukawa couplings and masses, see, e.g., Eq.~(\ref{eq:Zcoups_smallY}).  Consequently, they can differ by orders of magnitude. When these two particles are produced together, it is therefore plausible that a single displaced vertex occurs, and  the other particle is sufficiently long-lived as to escape the detector. Alternately, a displaced vertex may occur in conjunction with a promptly-decaying particle. To get a feeling for when displaced decays are important, we give expressions for $c \tau$ assuming that three-body decays dominate. The lifetime of the $\chi^0_{2}$ state is approximately given as
\begin{equation}
c \tau_{\chi_2} \approx 0.2  {\; \rm cm} \left( \frac{10^{-4}}{y_{+}}\right)^2  \left(\frac{10 \, {\rm GeV}}{\Delta m} \right)^3.
\end{equation}
The  scaling $\propto (\Delta m)^3$ reflects a cancelation of two powers in the typical three-body scaling $(\Delta m )^5$  with factors from the couplings $\propto y_{+}/\Delta m$. 
The $\chi^0_{3}$ lifetime, when its decay amplitude is dominated by an intermediate off-shell $Z$-boson, is approximately
\begin{equation}\label{eq:cTau3Z}
    c \tau_{\chi_3}^{Z} \approx 20 {\; \rm cm} \left( \frac{10^{-4}}{y_{-}}\right)^2 \left( \frac{M_{D}}{100 \, {\rm GeV}}\right)^2 \left(\frac{10 \, {\rm GeV}}{\Delta m} \right)^5.
\end{equation}
This expression is valid for $y_{2}/y_{1}$ not too close to 1, see below.

To get a feel for the lifetimes, in Fig.~\ref{fig:ExperimentalSignature} (left panel), we display contours of $c \tau$ for the neutral doublet states as a function of $y_{2}/y_{1}$ and $y_{1}$.  
Here, $M_D=250$ GeV, and the value
$M_S \approx 234$ GeV reproduces the observed relic abundance for the values of $y_{1}$ shown.  These Yukawa couplings correspond to the plateau. In general $\chi^0_3$ has a much longer lifetime than $\chi^0_2$ except in the vicinity of $y_2/y_1=-1$. This opens the possibility of signatures in which one state decays promptly while the other produces a displaced vertex. Over the range of Yukawa couplings shown, the model lies almost entirely in the co-annihilation regime, where the mass splitting is essentially independent of the Yukawa couplings. In this regime, the decay widths  $\Gamma_{\chi_{i}} \propto y_{1}^2 f_{i}\left( (y_{2} /y_{1})^2 \right)$.   The contours reflect this simple scaling.
  The right panel of Fig.~\ref{fig:ExperimentalSignature} breaks the regions of the contour plot down by the possible experimental signatures. 
There, we classify $\chi^0_2$ and $\chi^0_3$ into one of three categories according to its decay length: prompt (P), defined by $c\tau<1$ mm, displaced (D), defined by 1 mm $<c\tau<$ 1 m, and detector stable (S), defined by $c\tau>1$ m. Each region is labeled by two letters; the first denotes the decay behavior of $\chi^0_2$ and the second that of $\chi^0_3$. For example, the region labeled PD corresponds to $\chi^0_2$ decaying promptly while $\chi^0_3$ produces a displaced vertex. Eight distinct regions are realized. 
We do not find any parameters for which $\chi^0_2$ is detector stable but $\chi^0_3$ decays promptly.

\begin{figure}
    \centering

    \begin{subfigure}[t]{0.49\linewidth}
        \centering
        \includegraphics[width=\linewidth]{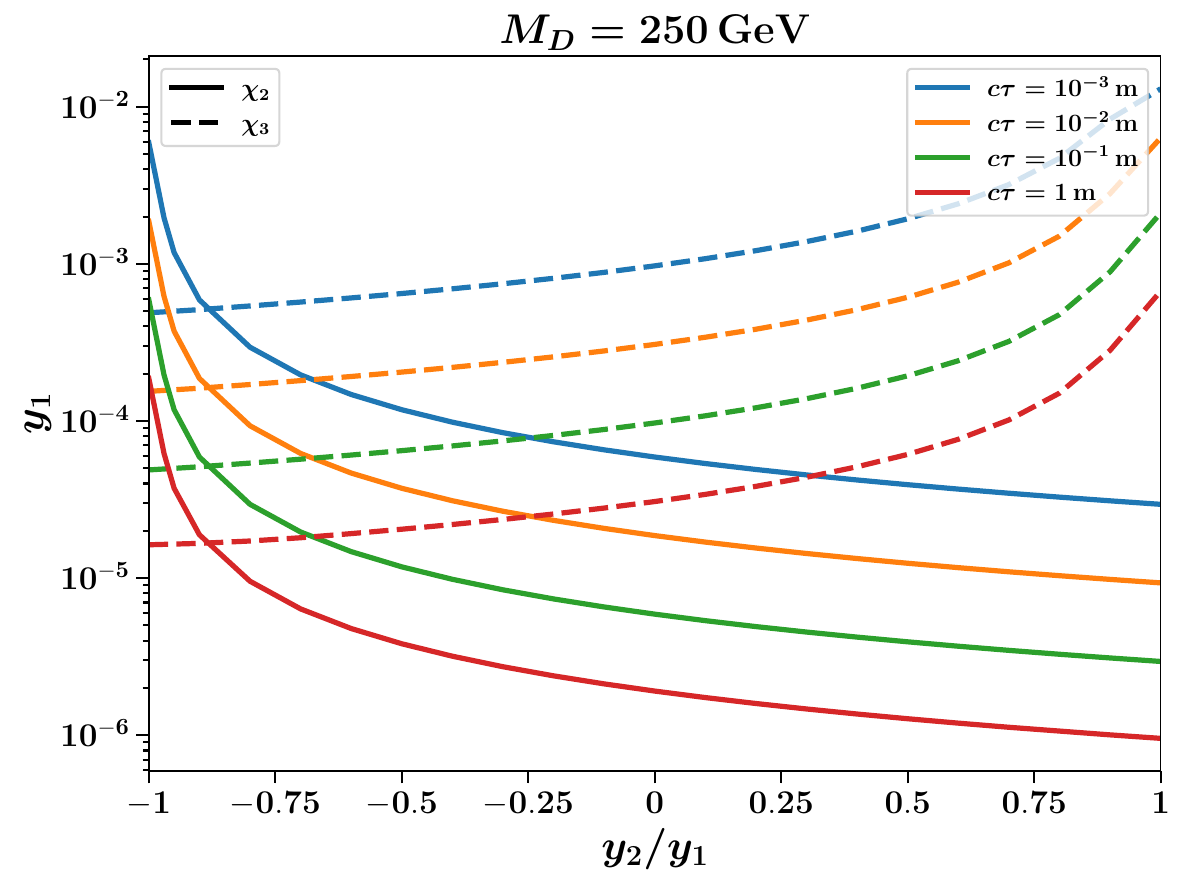}
    \end{subfigure}
    \hfill
    \begin{subfigure}[t]{0.47\linewidth}
        \centering
        \includegraphics[width=\linewidth]{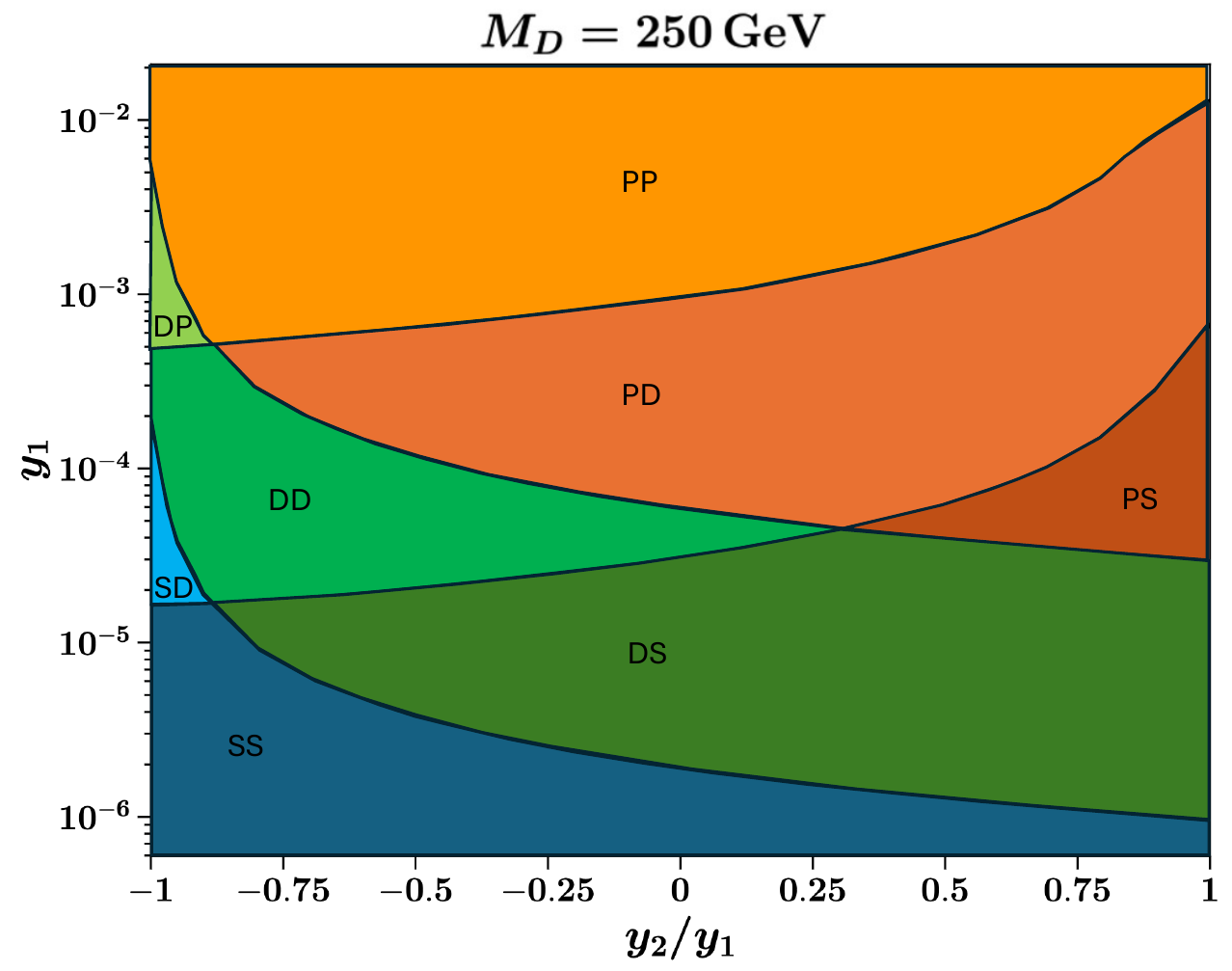}
    \end{subfigure}

    \caption{Left: Contours of $c\tau$ from 1 mm to 1 m for $\chi^0_2$ and $\chi^0_3$ in the $y_{2}/y_{1}$ vs. $y_{1}$ plane. Here the mass of the doublet is fixed at $M_D=250$ GeV while $M_S=234$ GeV  to reproduce the observed relic density. Right: Visualization of the same parameter space, classifying the decay behavior of $\chi^0_2$ and $\chi^0_3$. Each state can decay promptly (P), defined by $c\tau<1~\mathrm{mm}$; produce a displaced vertex (D), corresponding to $1~\mathrm{mm}<c\tau<1~\mathrm{m}$; or be detector-stable (S), with $c\tau>1~\mathrm{m}$. The eight regions are labeled by two letters, with the first indicating the decay behavior of $\chi^0_2$ and the second $\chi^0_3$.}
    \label{fig:ExperimentalSignature}
\end{figure}

Next, we comment on the fate of the charged state in our model.  The precise splitting between the charged-doublet state and the neutral-doublet state was computed (in the closely related supersymmetric context) in  \cite{Ibe:2023dcu}, building on the work of \cite{Thomas:1998wy}.  This is an electroweak symmetry breaking effect, and for large $M_{D}\gg M_Z$ and small Yukawa couplings, the one-loop value of the splitting $\delta_{+0}\equiv m_{\chi^{+}} - m_{\chi_2}$ asymptotes to $\frac{\alpha}{2} M_{Z}$.  Numerically, the asymptotic value is $\delta_{+0}  \rightarrow 353$ MeV \cite{Ibe:2023dcu}, which includes an $\approx 2$ MeV correction to the leading one-loop result.  This mass splitting readily admits a decay to a charged pion with partial width given by 
\begin{equation}\label{eq:Cha_to_chi2_pion}
\Gamma(\chi^{+} \rightarrow \chi^0_{2} \pi^{+}) =\frac{G_F^2}{2\pi}\cos^2{\theta_c}f_\pi^2\delta_{+0}^3\sqrt{1-\left(\frac{m_\pi}{\delta_{+0}}\right)^2},
\end{equation}
with $G_{F}$ the Fermi constant and $f_\pi\approx130$ MeV the pion decay constant.
The partial width $\Gamma(\chi^{+} \rightarrow \chi^0_{3} \pi^{+})$ is identical (once Yukawa couplings are small enough that $\chi^0_{2}$ and $\chi^0_{3}$ are degenerate on the scale of the charged/neutral mass splitting).
If these decays dominate, the lifetime is of order $\tau \sim 0.05$ ns, and the final state is simply a soft pion produced after a short disappearing track.

With an understanding of the range of lifetimes that are available for the doublet states, we now turn to a review of current and potential searches. No search represents an exact match for our model, and we plan a detailed recast in future work.  Here our goal is  to draw attention to relevant searches and provide a rough estimate for the current and future sensitivity to the model. We first outline the channels involving three-body decays.
If Yukawa couplings are sufficiently large, all decays are prompt.  Standard searches motivated by electroweakinos would apply.  For related work, see \cite{Calibbi:2015nha, Freitas:2015hsa} and experimental searches \cite{ATLAS:2024lda, CMS:2024gyw, ATLAS:2024qxh}.  While no search precisely reproduces this model, the phenomenology is perhaps most similar to the Bino-Wino benchmark of \cite{CMS:2024gyw}, although the production cross section in our case is  smaller than the one for the corresponding cross section for the Wino for a given mass.  Correcting for this difference in cross section, we can estimate that dark matter masses up to $\sim$ 200 GeV would be excluded at present. 

Experiments at the LHC have also performed a variety of searches for displaced decays.
 A  phenomenology somewhat related to the current model (where a dark sector had displaced  decays of the form $\chi^{+} \rightarrow W^{\ast} \chi^{0}$) was discussed in \cite{Blekman:2020hwr}, and was later searched for by ATLAS  \cite{ATLAS:2024vnc}.  However, our model motivates decays not via an off-shell $W^{\ast}$, but rather a $Z^{*}$ (recall the charged states typically decay relatively quickly to a soft pion). The CMS experiment searched \cite{CMS:2025qkk} for a similar topology in a Bino/Wino simplified model, motivated in part by \cite{Nagata:2015pra}.  The sensitivity of this search is highly dependent on the $c \tau$ of the heavy neutral state.  Again, when compared to the Bino-Wino benchmark of the search, the present model has a reduced production cross section and differs in that here the chargino decays to a pion with a short disappearing track.  The sensitivity of the CMS search peaks around $c \tau \approx 10 {\rm \, mm}$. 
A naive re-interpretation of the bounds from this search suggests sensitivity  to doublet masses approaching $\sim 300$ GeV for $c \tau$ in this range.  The search sensitivity drops rapidly at other lifetimes.
A search was very recently conducted by ATLAS \cite{ATLAS:2026riv} that looked for similar displaced di-lepton topologies, but with a focus on much harder 
leptons, and this search did not allow for the possibility of the off-shell $Z$ topology, nor did it take advantage of the possible large missing energy that might occur in this signal.

A similar search by CMS \cite{CMS:2023bay} looked for an inelastic dark matter model where dark matter $\chi^0_{1}$ was produced in association with a dark sector partner $\chi^0_{2}$, which subsequently decayed to the dark matter particle via  emission of an off-shell dark photon $\chi^0_{2} \rightarrow A'^{\ast} \chi^0_{1} \rightarrow \mu^{+}\mu^{-} \chi^0_{1}$. This model allows for the possibility of a displaced muon pair, and so took advantage of a displaced vertex reconstruction algorithm.  This analysis places bounds on $\sigma_{\chi^0_{2} \chi^0_{1}} \times BR (\chi^0_{2} \rightarrow \chi^0_{1} \mu^{+} \mu^{-})$, with limits in the 10$^{-1}$--10$^{-2}$ pb range for similar, but not identical, kinematics.  The development of a dedicated trigger to search for displaced decays (perhaps in conjunction with disappearing tracks, see below), along the lines of the CMS displaced trigger \cite{CMS:2026blj} could be of great interest for the phenomenology discussed here. 

Above, we discussed the possibility that the  $\chi^0_{3}$ state can have a significant branching ratio to $\chi^0_{1} + \gamma$. If it were to have a long lifetime, it is possible that a search for a non-pointing photon  along the lines often to search for gauge-mediated supersymmetry breaking \cite{ATLAS:2014kbb, CMS:2019zxa} for non-pointing  photons could be employed.  However, because $\chi^0_{3}$ pair production is suppressed, we do not expect di-photon final states, so a di-photon trigger would not be efficient. Sensitivity non-pointing photons is typically best for lifetimes around 1 ns \cite{ATLAS:2014kbb, CMS:2019zxa}.

In the freeze-in and very-low-Yukawa-coupling regime, the presence of the singlet dark matter is immaterial for collider signals.   Its couplings to the SM are too feeble for it to be produced, and decays from the doublet sector to the singlet are too slow to matter.  For this reason, studies that consider a pure doublet can essentially be ported directly. Essentially, $\chi^0_{2}$ and $\chi^0_{3}$  appear as missing energy, while the $\chi^{\pm}$ give rise to a disappearing track that produces a soft track when the $\chi^{\pm}$ decay to the other doublet states.

The $\chi^{\pm}$ lifetime is sufficiently short as to make searches via disappearing tracks challenging \cite{Halverson:2014nwa}. Nevertheless, the LHC has been able to extend the reach beyond that of LEP, and the result from 137 fb$^{-1}$ \cite{ATLAS:2026jhep} excludes $M_{D} < 225$ GeV at 95\% CL.  At this mass, the lifetime is approximately $\tau_{+} \simeq 0.04$ ns. 
The ATLAS and CMS collaboration released a study  \cite{ATLAS:2018jjf} in 2018 indicating that the high-luminosity LHC is expected to extend these searches to 270 GeV  at 95\%  using 3000 fb$^{-1}$.  It has also been proposed \cite{Fukuda:2019kbp} that the soft tracks that arise from the $\chi^{\pm}$ decays might be useful.  They projected that searches using these soft tracks could probe doublet masses up to approximately 210 GeV with 3000 fb$^{-1}$ of data.  Monojet searches at the LHC that effectively search for the doublet sector are also of potential use, and they generally extend to higher masses.  For example, with 3000 fb$^{-1}$ of data, the HL-LHC should be able to exclude models at the 95\%  level up to masses of approximately 375 GeV \cite{Belyaev:2018ext,Belyaev:2020wok}. However, a monojet signal in isolation would not give any information about the presence of the two closely spaced-states that are indicative of the doublet sector.

A future muon collider could be an especially powerful probe of this model.  Not only could searches for disappearing tracks be done, see e.g., \cite{Han:2022ubw, Capdevilla:2021fmj}, 
recently, Ref.~\cite{Capdevilla:2024bwt} studied the potential for using the soft tracks at a 3 TeV MuC.  They found these searches are especially powerful reach for the lifetimes of interest, and should effectively be able to probe the entire mass range of interest, with a 5$\sigma$ discovery possible up to approximately 1250 GeV.  If this strategy proves feasible, it could definitively determine whether the doublet structure of this model is present.  Unfortunately, there would be no way to observe the singlet-like dark matter directly, however.

Finally, we comment on the potential power of precision electroweak (PEW) measurements. A very recent analysis \cite{Hamaguchi:2026iry} highlights the potentially important role that PEW measurements at a future $e^{+} e^{-}$ machine could play in probing this scenario.  As noted in \cite{Marandella:2005wc,Martin:2004id}, a pure doublet makes contributions to the $W$ and $Y$ oblique parameters, which in the heavy doublet limit are given by
\begin{equation}
W \simeq 2.3 \times 10^{-4} \left(\frac{100 {\rm \, GeV}}{M_{D}}\right)^2, \qquad  Y \simeq \tan^2 \theta_{
W} \, W.
\end{equation}
This might produce a $2\sigma$ deviation up to masses $M_{D}$ of 500 GeV at FCC-ee \cite{Hamaguchi:2026iry}.


\section{Conclusion}
The Singlet-Doublet dark matter model is notable because it relies only on known force carriers to realize the observed dark matter relic abundance.  We have examined this model in a variety of regimes, including co-annihilation, co-scattering, freeze-in, and the SuperWIMP mechanism.  In all these regimes, it is possible to realize a thermal history that successfully reproduces the observed dark matter abundance.  When Yukawa couplings are small, and the dark matter is primarily singlet, direct and indirect detection in this model is very strongly suppressed.  However, the presence of doublet states presents a target for current and future colliders.  We have outlined a variety of searches that apply for different values of the Yukawa coupling.  Of particular interest are searches with long-lived particles and displaced vertices, sometimes produced in conjunction with disappearing and/or soft tracks.  These signatures might present a target for long-lived particle triggers at the LHC.

\begin{acknowledgments}
We would like to thank C.~Herwig for useful conversations and E.~Petrosky for collaboration in the early stages of this work. AP is supported by
the Department of Energy under grant number DE-SC0007859. 
PNB is supported by the European Union - Next Generation EU through the PRIN2022 Grant n. 202289JEW4.
This research was supported in part through computational resources and services provided by Advanced Research Computing (ARC), a division of Information and Technology Services (ITS) at the University of Michigan, Ann Arbor. 
\end{acknowledgments}

\bibliographystyle{apsrev4-1}
\bibliography{ref}

\end{document}